\documentclass[reprint,nofootinbib,amsmath,amssymb,aps,prmaterials]{revtex4-2}

\usepackage{graphicx}
\usepackage[dvipsnames]{xcolor}
\usepackage[colorlinks=true,
    linkcolor=blue,
    filecolor=blue,
    citecolor=blue,      
    urlcolor=blue,]{hyperref}

\usepackage{prettyref}
\newcommand{\pref}[1]{\prettyref{#1}}
\newrefformat{fig}{Fig.~\ref{#1}}
\newrefformat{tab}{Table~\ref{#1}}
\newrefformat{sec}{Sec.~\ref{#1}}
\newrefformat{ssec}{Sec.~\ref{#1}}
\newrefformat{app}{App.~\ref{#1}}
\newrefformat{eqn}{Eq.~(\ref{#1})}

\usepackage{textcomp}

\begin{document}

\title{Oxygen vacancies in vanadium dioxide: A DFT+\textit{V} study}

\author{Oskar Leibnitz}
\author{Peter Mlkvik}
\email{peter.mlkvik@mat.ethz.ch}
\author{Nicola A. Spaldin}
\author{Claude Ederer}
\affiliation{Materials Theory, Department of Materials, ETH Z\"{u}rich, Wolfgang-Pauli-Strasse 27, 8093 Z\"{u}rich, Switzerland}

\date{\today}

\begin{abstract}
We present a density-functional theory study of the effects of oxygen vacancies on the structural and electronic properties of vanadium dioxide (VO$_2$). Our motivation is the reported suppression of the metal-insulator transition by oxygen vacancies and the lack of a clear consensus on its origin. We use the DFT$+V$ method with a static intersite vanadium-vanadium interaction term, $V$, to calculate the properties of the oxygen-deficient metallic rutile and insulating monoclinic M1 phases of VO$_2$ on the same footing. We find that oxygen vacancies induce local distortions in the M1 phase, but do not destroy the dimerization usually associated with the insulating behavior. In spite of this, we find that the M1 phase becomes metallic as a result of the partial filling of the conduction band due to a rigid-band-like doping effect.
\end{abstract}

\maketitle


\section{Introduction}

The metal-insulator transition (MIT) in vanadium dioxide (VO$_2$) has been proposed for a variety of applications, from memristive devices~\cite{Yi_et_al:2018, Yuan_et_al:2023}, ultra-fast electronics~\cite{Yang/Ko/Ramanathan:2011, Zhou_et_al:2013}, to energy-efficient windows~\cite{Wang_et_al:2016, Cui_et_al:2018}. In order to fully realize the application potential of VO$_2$, the MIT temperature, $T_c$, as well as the transition hysteresis, and order of magnitude change in resistivity, should ideally be reliably tuneable. Besides epitaxial strain~\cite{Cao_et_al:2010, Atkin_et_al:2012, Aetukuri_et_al:2013, Makarevich_et_al:2021}, and doping~\cite{Strelcov_et_al:2012, Xue/Yin:2022, Krammer_et_al:2017, Mlkvik/Ederer/Spaldin:2022}, oxygen vacancies (O$_V$'s), one of the most frequent point defects, have been studied extensively for their potential to mediate such tuning. In this work, we computationally investigate the effects of oxygen vacancies on the main VO$_2$ phases.

\begin{figure}
    \centering
    \includegraphics[width=0.8\linewidth]{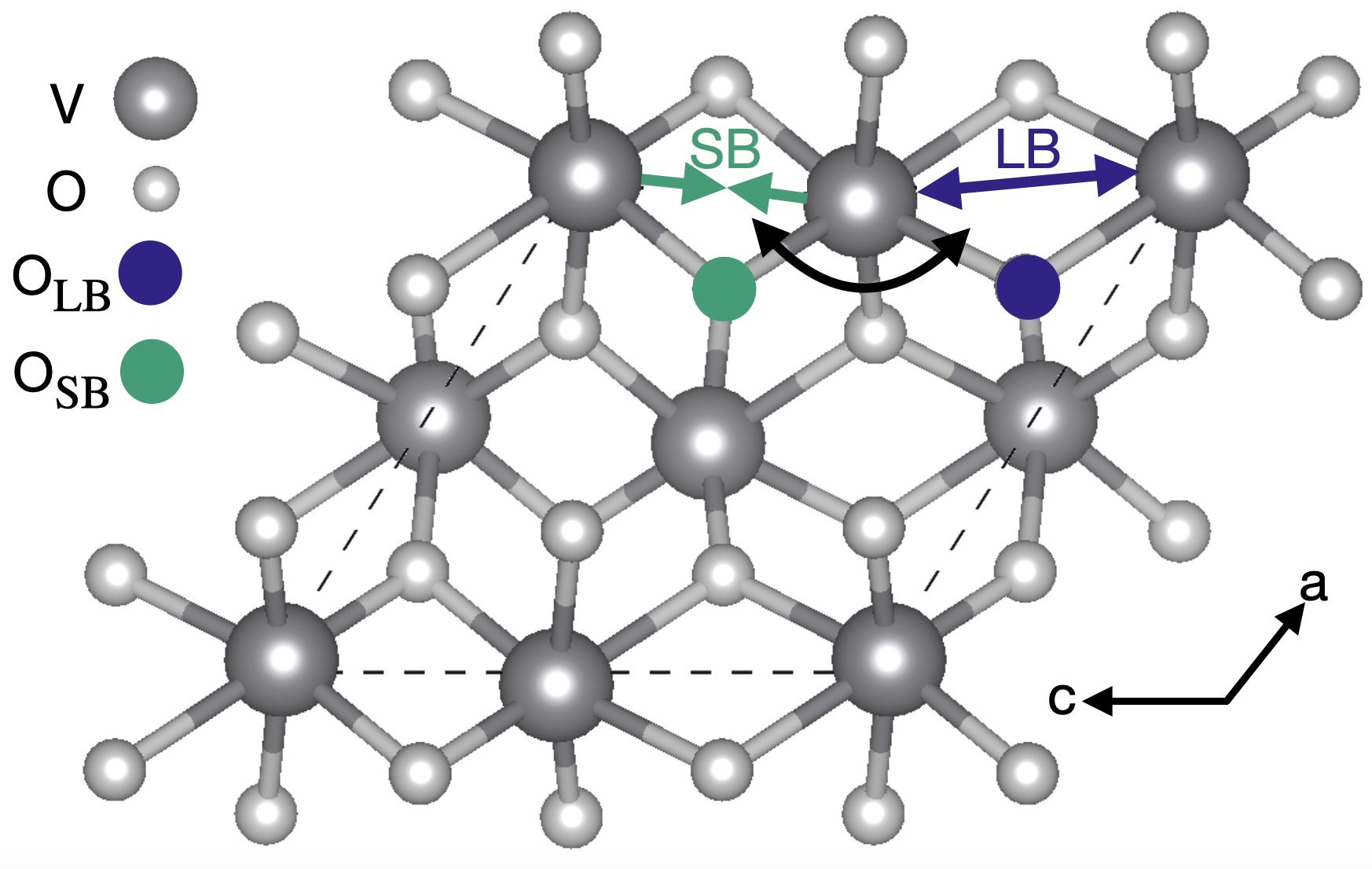}
    \caption{Crystal structure of the M1 phase in the M1 primitive cell (dashed line). Green and blue arrows indicate the short and long bonds (SB and LB, respectively) along the $c$ direction. V (O) atoms shown in (light) gray. Inequivalent oxygen sites in the M1 structure (O$_\text{SB}$ and O$_\text{LB}$) are indicated in green and blue, respectively.}
    \label{fig:phases}
\end{figure}

The MIT in VO$_2$ is accompanied by a structural transition, from a high-temperature rutile (R) phase to a low-temperature monoclinic (M1) phase, at $T_c=340$\,K~\cite{Morin:1959, Eyert:2002a}. The transition involves V--V dimerization along the tetragonal $c$ axis coupled to lateral displacements, causing a doubling of the unit cell, resembling a Peierls instability. In the R phase, $3d^1$ V$^{4+}$ ions situated in distorted VO$_6$ octahedra have partially filled $a_{1g}$ and $e_g^\pi$ states, resulting from a splitting of the $t_{2g}$ manifold. In the dimerized M1 phase (\pref{fig:phases}), the $a_{1g}$ band is further split into occupied bonding and unoccupied antibonding states~\cite{Goodenough:1971}, while the $e_g^\pi$ levels shift upward, resulting in a narrow insulating gap. Besides these Peierls effects, Mott-Hubbard correlation effects are needed for a full description of the physics of VO$_2$~\cite{Pouget_et_al:1974, Bianconi:1982, Zylbersztejn/Mott:1975, Rice/Launois/Pouget:1994, Lee_et_al:2018, Weber_et_al:2012, Gatti_et_al:2007, Brito_et_al:2016, Najera_et_al:2017}, and a dual Peierls-Mott mechanism appears to provide the most appropriate description of the MIT in VO$_2$~\cite{Biermann_et_al:2005, Tomczak/Biermann:2007, Grandi/Amaricci/Fabrizio:2020, Mlkvik_et_al:2024}.

Numerous studies consistently show that increasing oxygen vacancy concentration suppresses the insulating state of VO$_2$. Oxygen vacancies lower the $T_c$ even when strain effects are excluded~\cite{Fan_et_al:2018, Fan_et_al:2013a, Jeong_et_al:2013, Basyooni_et_al:2022, Sim_et_al:2024}. Field-switching experiments reveal a corresponding decrease in the critical voltage~\cite{Zhang_et_al:2017a}, and O$_V$'s can modify or even eliminate hysteresis~\cite{Zhang_et_al:2017, Lu_et_al:2020a}. Isotope-labeled oxygen studies directly confirm that these effects arise from oxygen vacancies~\cite{Jeong_et_al:2013}. Quantifying the vacancy concentration remains difficult, with most studies offering only qualitative comparisons, such as different vacuum annealing times~\cite{Chen_et_al:2016a}. X-ray photoemission spectroscopy provides rare quantitative estimates, reporting $\delta = 0.07\, - \,0.20$~\cite{Xu_et_al:2016,Zhang_et_al:2017a} in VO$_{2-\delta}$.

Despite this phenomenological consensus, the literature diverges on the nature of the physics underlying the change in $T_c$, with two main interpretations emerging. First, several studies suggest that oxygen vacancies stabilize the R structure below the $T_c$ of the stoichiometric phase. Xu~\textit{et al.}~\cite{Xu_et_al:2016} observe a softening of Raman modes associated with the M1 phase and a lowering of the structural transition temperature in thin films of VO$_{1.92}$. The proposed explanation is a weakening of the V--O hybridization and the V--V dimerization by the electrons donated by the vacancy. Kim \textit{et al.}~\cite{Kim_et_al:2014} similarly observe a reduction in the structural transition temperature, but attribute it to a weakening of the V--V dimers by oxygen-vacancy-induced compressive strain. Zhang~\textit{et al.}~\cite{Zhang_et_al:2017a} report Raman mode softening at a temperature below $T_c$ and link it to a stabilization of the R phase due to the dopant electrons. Matsuda~\textit{et al.}~\cite{Matsuda_et_al:2025} show spectroscopic evidence of a confinement of the V--V pairing to nanodomains within an undimerized R-like VO$_2$ structure at room temperature, consistent with partial rutile phase stabilization due to oxygen deficiency. Guo~\textit{et al.}~\cite{Guo_et_al:2024} report a decrease in $T_{\text{c}}$ with Ar-plasma irradiation in VO$_2$ nanobeams and link it to oxygen vacancy induced stabilization of the R phase; however, no experimental phase characterization is performed. Passarello~\textit{et al.}~\cite{Passarello_et_al:2016} demonstrate that ionic liquid gating induces oxygen vacancy formation and stabilizes the R structure down to 270~K. The retention of R symmetry is confirmed spectroscopically and attributed to a combination of vacancy-driven lattice expansion and epitaxial clamping of the in-plane lattice constants by the TiO$_2$ substrate used. 

The second interpretation common in literature is the emergence of a metallic M1 phase. This phase has been observed through pump-probe spectroscopy and reported in non-equilibrium studies on photoexcited VO$_2$~\cite{Morrison_et_al:2014, Wegkamp_et_al:2014}. A variety of other studies support the stabilization of such a metallic M1 phase in oxygen deficient VO$_2$. Li~\textit{et al.}~\cite{Li_et_al:2017} observe a vacancy-induced metallic M1 phase in VO$_2$ nanobeams and associate it with orbital-selective carrier doping. Schrecongost~\textit{et al.}~\cite{Schrecongost_et_al:2019, Schrecongost_et_al:2020, Schrecongost_et_al:2022} demonstrate reconfigurable nanoplasmonic devices based on metallic M1 phases patterned via oxygen vacancy or free carrier injection using atomic force microscopy, attributing the metallicity to the associated carrier injection. The monoclinic nature of the observed phase is confirmed using micro x-ray diffraction and electron diffraction. Vacuum annealing experiments reveal a metallic M1 phase, confirmed by hardening of Raman-active V--V modes, dimer stabilization, and robust room-temperature metallicity, and attribute it to changes in electronic correlations and a shortening of V--V bonds~\cite{Chen_et_al:2016a}. 

The effect of oxygen vacancies has also been studied using first-principles methods. Several DFT$+U$ studies~\cite{Fan_et_al:2018, Chen_et_al:2016a, Liang_et_al:2016, Appavoo_et_al:2012} report a metallic M1 phase as the ground state of VO$_2$, even at low vacancy concentrations, though their interpretations vary. Guo~\textit{et al.}~\cite{Guo_et_al:2024} find partial filling of the conduction band in an oxygen deficient M1 structure and deduce a stabilization of the R phase from this result, although they do not explicitly calculate the non-stoichiometric R structures. Chen~\textit{et al.}~\cite{Chen_et_al:2016b} also only study the oxygen-deficient M1 phase, finding weaker V--V dimerization from which the authors infer a total de-dimerization into an R phase under larger oxygen concentrations. Lee~\textit{et al.}~\cite{Lee_et_al:2018} consider hetero-structures of oxygen-deficient and stoichiometric VO$_2$ and construct a two-dimensional energy phase space describing the electronic correlation and the structural V--V dimerization. Their model exhibits a global energy minimum at high electronic correlation and large dimerization (insulating M1), but also a local minimum at low correlation and large dimerization, which they identify with the metallic M1 phase stemming from the oxygen deficient state stabilized by the presence of the underlying stoichiometric VO$_2$. Ganesh~\textit{et~al.}~\cite{Ganesh_et_al:2020} use dynamical mean-field theory and diffusion Monte Carlo to argue for the suppression of the MIT due to a vacancy-induced reduction of the partial $a_{1g}$ orbital polarization in the R structure. This reduces the dimerization tendency, hence stabilizing the R phase. Their findings support the existence of a quasi-percolation threshold for vacancies above which the MIT is fully suppressed, linking it to experimental observations of a critical vacancy concentration~\cite{Zhang_et_al:2017}. In addition, the authors calculate the optical band gap in M1 VO$_{2-\delta}$ and observe its reduction down to zero depending on the vacancy position, effectively observing a metallic M1 phase.
 
In summary, while the literature consistently describes a suppression of the MIT in oxygen-deficient VO$_2$, decisively different interpretations of the underlying physical effects exist. Computationally, no existing study adequately captures both the structural and electronic effects across the R and M1 phases within a unified framework. A comprehensive approach is therefore highly desirable to make a conclusive statement about the mechanism of MIT suppression in VO$_{2-\delta}$. Recently, in Ref.~\cite{Haas_et_al:2024}, we have demonstrated that incorporating a static intersite interaction term within the DFT$+V$ method enables efficient, accurate, and unbiased treatment of both the M1 and R phases on the same footing. Accordingly, in this work, we study the phase stability and electronic structure in VO$_{2-\delta}$ using the DFT$+V$ method. 


\section{Computational details}

In our calculations, we use different periodic cells, all based on the unit cell of the low-symmetry M1 structure, i.e., a doubled primitive cell of the high-symmetry R structure (\pref{fig:phases}). We use the experimental lattice parameters of the R phase~\cite{McWhan_et_al:1974}, and neglect the monoclinic strain and the expansion of the $c$ lattice parameter in the M1 phase relative to R, which have a negligible effect on the resultant electronic structure~\cite{Haas_et_al:2024}. 

In this basic unit cell, a single oxygen vacancy results in VO$_{2-\delta}$ with $\delta = 0.25$, i.e., defect concentration of 12.5\%, and one oxygen vacancy appearing every two vanadium atoms along $c$ under periodic boundary conditions. To decrease the O$_V$ periodicity, we construct a $2\times 2 \times 2$ supercell (SC) of the stoichiometric primitive M1 unit cell, so that a single O$_V$ corresponds to $\delta \approx 0.03$ and a defect concentration of 1.5625\%, with one O$_V$ appearing every four vanadium atoms along $c$ under periodic boundary conditions. We additionally study a $1 \times 1 \times 5$ supercell elongated along the $c$ direction, with $\delta = 0.05$, resulting in a larger O$_V$ spacing along the V chains.

While all oxygen sites are symmetry equivalent in the R phase, the dimerization breaks this symmetry, leading to two distinct oxygen sites in the M1 phase. We label the sites adjacent to the short dimerized bond (SB) and long non-dimerized bond (LB) as O$_\text{SB}$ and O$_\text{LB}$, respectively (corresponding to OI and OII from Ref.~\cite{Ganesh_et_al:2020}), and we examine three different oxygen-deficient structures, R-O, M1-O$_\text{SB}$, and M1-O$_\text{LB}$. 

All DFT$+V$ calculations are performed using the \textsc{quantum espresso} (QE 7.3 \& 7.4) package~\cite{Giannozzi_et_al:2009, Giannozzi_et_al:2017, Giannozzi_et_al:2020}. The exchange-correlation energy is approximated using the Perdew-Burke-Ernzerhof (PBE) generalized gradient functional~\cite{Perdew/Burke/Ernzerhof:1996, Perdew/Yue:1986}. Ultrasoft pseudopotentials, which include the V 3$s$ and V 3$p$ semicore states in the valence manifold are sourced from the GBRV library~\cite{Garrity_et_al:2014}. Convergence tests establish 70~Ry as a suitable kinetic energy cutoff for the plane-wave basis set describing the Kohn-Sham wavefunctions and the kinetic energy cutoff for the charge density is set to 12\,$\times$\,70~Ry. A $\Gamma$-centered $6 \times 6 \times 8$ $\mathbf{k}$-point grid is used for the primitive cell and scaled down according to the reduction in Brillouin zone size in supercell calculations. The energy is converged to a tolerance of $5 \times10^{-8}$ eV between iterations. Electronic occupations are treated using Gaussian smearing with a width of $ 10^{-3} $ Ry to facilitate convergence while maintaining an effectively insulating occupation profile. Atomic positions are relaxed to a force tolerance of $ 10^{-3}~\mathrm{Ry}/\mathrm{bohr} $ using the BFGS algorithm. 

Following Ref.~\cite{Haas_et_al:2024} and using the implementation of Ref.~\cite{Timrov/Marzari/Cococcioni:2018} based on Ref.~\cite{LeiriaCampoJr/Cococcioni:2010}, we use DFT$+V$ as a static intersite correlation correction term in all calculations throughout this work. We treat the intersite interaction parameter $V$ as an empirical quantity modifying the corresponding effective hopping amplitudes. We set its value to 1\,eV, and apply it to 3$d$ projectors on all neighboring V--V pairs along the $c$ direction. The energy correction due to the intersite interaction, $V$, both depends on and modifies the intersite occupation, $n_\times$, where usually the trace (indicative of intersite same-orbital interactions) captures the majority of the intersite occupation. 

We construct maximally-localized Wannier functions~\cite{Marzari/Vanderbilt:1997} using the \textsc{wannier90} code~\cite{Pizzi_et_al:2020}. To construct the localized orbitals most relevant for understanding the MIT behavior, we select a subspace of V-$t_{2g}$-dominated bands located around the Fermi level. There are twelve $t_{2g}$ bands corresponding to three states for each of the four vanadium atoms in the stoichiometric primitive cell [see the energy range in \pref{fig:VacbandID}(b, c)]. Hence we initialize one of each $t_{2g}$-type projectors at each vanadium site, as well as including an additional $s$-type projector at each vacancy site (for motivation, see the discussion later and Refs.~\cite{Souto-Casares/Spaldin/Ederer:2019, Souto-Casares/Spaldin/Ederer:2021}).

The \textsc{banduppy} package~\cite{Medeiros_et_al:2015, Medeiros/Stafstrom/Bjork:2014, Iraola_et_al:2022} is used to compute unfolded band structures from supercell wavefunctions, using the unfolding formalism~\cite{Boykin/Klimeck:2005, Popescu/Zunger:2012}. Here, the irreducible set of supercell $\textbf{k}$ points is mapped, \textit{unfolded}, onto a larger set of primitive cell $\textbf{k}$ points. A spectral weight, $A(\textbf{k},E)$, a momentum-resolved primitive-cell representation of the electronic structure at each $\textbf{k}$ point, is then computed directly from the supercell plane-wave data without requiring a separate primitive-cell calculation.


\section{Results and Discussion}

We start by considering a primitive cell of M1-O$_\text{SB}$ VO$_{2-\delta}$ with a single vacancy. Although this cell describes a defect arrangement with relatively short periodicity (O$_V$ every two vanadium atoms along $c$), it nevertheless serves as a first demonstration of the behavior of O$_V$ in VO$_{2-\delta}$.

\begin{figure}
    \centering
    \includegraphics[width=1\linewidth]{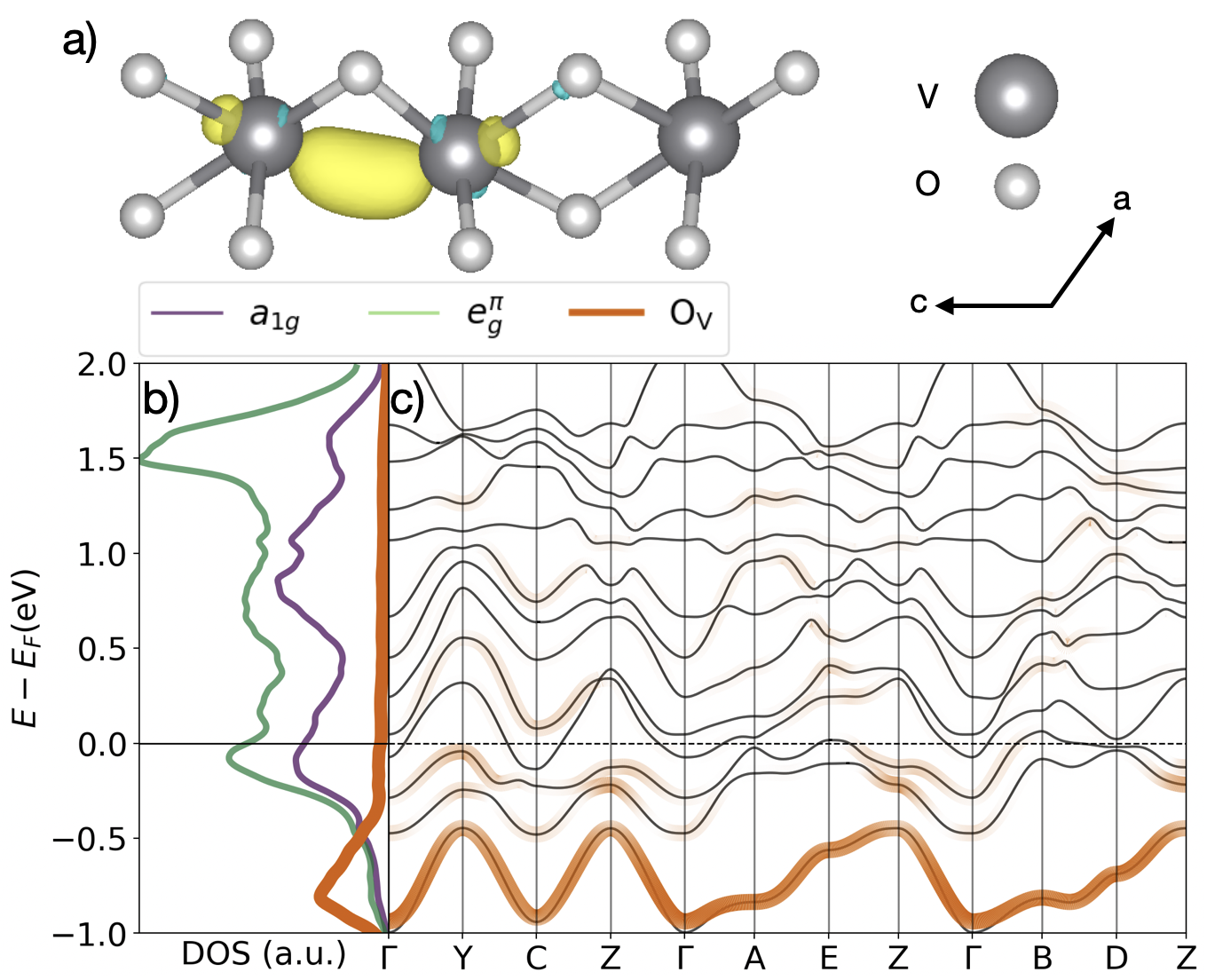}
    \caption{ (a) Detail of the primitive cell M1-O$_\text{SB}$ vacancy WF. Yellow and blue colors indicate positive and negative phase, respectively, V (O) atoms shown in (light) gray. (b) PDOS of the primitive cell with purple (green) colored lines indicating the $a_{1g}$ ($e_g^\pi$) WF. O$_V$ shown in orange. (c) Band structure of the primitive cell M1-O$_\text{SB}$ with the O$_V$ contribution shown in orange.}
    \label{fig:VacbandID}
\end{figure}

In \pref{fig:VacbandID}(b, c), we show the projected density of states (PDOS) and the band structure of the relaxed primitive cell M1-O$_\text{SB}$ structure. We observe an emergence of an additional band within the energy range of the $t_{2g}$ manifold (centered around $-0.7$\,eV) compared to the stoichiometric structure~[\pref{fig:pristine}(a, b)]. This feature appears consistently across all oxygen-deficient configurations presented in this work. In fact, if we initialize a Wannier function (WF) at the vacancy position, we find that this orbital maps almost completely onto the new-found band [see the vacancy contribution in orange in \pref{fig:VacbandID}(b, c)], and cannot be directly attributed to any other manifold in the full band structure. The resultant WF is highly localized in space (with spread comparable to those of the $t_{2g}$ WFs), stretching between the neighboring vanadium atoms across the SB [\pref{fig:VacbandID}(a)]. Notably, the vacancy state is also strongly localized in energy.

We note that no clear $a_{1g}$ bonding-antibonding splitting is visible in the PDOS in \pref{fig:VacbandID}(b), and there is also a relatively large occupation of the $e_g^\pi$ orbitals, unlike the stoichiometric case shown in \pref{fig:pristine}(a, b). However, since this small cell is not very representative, we do not analyze this any further. We mainly include these results to introduce the vacancy WF, which is most prominent in the small cell with high vacancy concentration. Due to the short periodicity resulting from the small size of the primitive cell, we focus the remainder of our analysis on larger supercells. 

\subsection{Fixed atomic positions}

We first consider unrelaxed supercells, where we remove an oxygen atom but otherwise keep the atomic positions fixed, to assess the purely electronic impact of an oxygen vacancy. 

In \pref{tab:energies}, we show the energy difference ($\Delta E$) between the different VO$_2$ and VO$_{2-\delta}$ phases and configurations. In the unrelaxed, fixed-geometry case, the R-O structure remains significantly disfavored (see middle rows in \pref{tab:energies}), with energy difference similar to stoichiometric VO$_2$ (top rows in \pref{tab:energies}). Both monoclinic configurations exhibit lower total energies, with the M1-O$_\text{LB}$ slightly favored over the M1-O$_\text{SB}$ structure.

\begin{figure}
    \centering
    \includegraphics[width=1\linewidth]{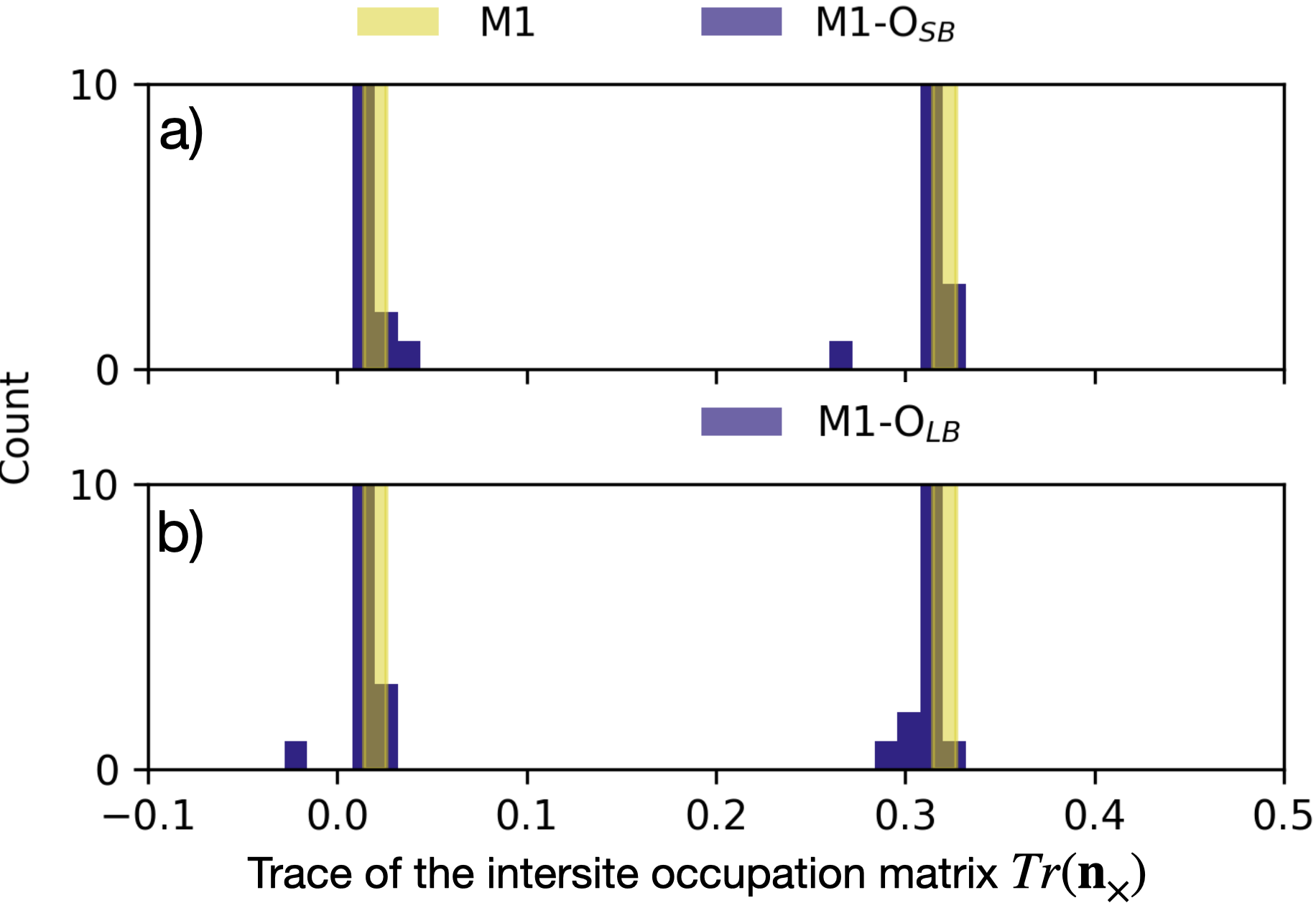}
    \caption{The trace of the intersite occupation matrix, $\text{Tr}(n_\times)$, for neighboring V--V pairs along $c$ in (a) the M1-O$_\text{SB}$ and (b) the M1-O$_\text{LB}$ structures (blue) compared to stoichiometric M1 values (yellow).}
    \label{fig:nxx}
\end{figure}

To better understand the effects of the vacancy, we investigate the electronic effects, and specifically the intersite occupation as a quantity representing the V--V bonding. In \pref{fig:nxx}(a, b), we show the trace of the intersite occupation matrix, $\text{Tr}(n_\times)$, for all neighboring V--V pairs along the $c$ direction in both stoichiometric VO$_2$ and VO$_{2-\delta}$ (yellow and blue lines, respectively). For M1 VO$_2$, $\text{Tr}(n_\times)$ yields two values, corresponding to short and long distances between vanadium atoms, and hence large and small values of the electronic overlap. 

Upon introduction of a vacancy, the distribution of $\text{Tr}(n_\times)$ values for both M1 configurations remain close to those of the stoichiometric reference values. In each case, there is only a single data point with significantly different $\text{Tr}(n_\times)$ corresponding to the vanadium pair adjacent to the vacancy, either on the short (M1-O$_\text{SB}$) or the long (M1-O$_\text{LB}$) bond between nearest neighbors [at 0.27 in \pref{fig:nxx}(a) and at $-0.02$ in \pref{fig:nxx}(b), respectively]. The reduction of this $\text{Tr}(n_\times)$ value in the M1-O$_\text{SB}$ configuration already indicates a suppression of hybridization between nearest-neighbor orbitals by the presence of the vacancy, even without structural relaxation. Notably, this reduction is rather small, highlighting that the dimerization in VO$_2$ largely stems from a direct V--V overlap rather than oxygen atoms mediating the V--V intersite interaction.

The unperturbed nature of the majority of V--V pairs except for those closest to the vacancy indicate the high localization of the vacancy effects in the M1 structures.

\begin{table}
    \centering
    \begin{tabular}{c  c  c}
        \hline\hline 
        Geometry & Structure & $\Delta E$ [meV/f.u.] \\ \hline
        Stoichiometric  & R       & 75.55    \\
        Stoichiometric  & M1      & 0    \\ \hline
        Unrelaxed & R-O     & 74.9 \\ 
        Unrelaxed & M1-O$_\text{SB}$   & 3.7  \\ 
        Unrelaxed & M1-O$_\text{LB}$  & 0  \\ \hline
        Relaxed   & R-O     & 32.27  \\ 
        Relaxed   & M1-O$_\text{SB}$   & 0.06   \\ 
        Relaxed   & M1-O$_\text{LB}$  & 0      \\ \hline\hline
    \end{tabular}
    \caption{Relative energies of the different structures of VO$_2$ and VO$_{2-\delta}$, for $\delta\approx 0.03$. $\Delta E = E - E_\text{min}$, with $E_\text{min}$ taken from the respective minimum energy structure.}
    \label{tab:energies}
\end{table}

\begin{figure*}[t]
    \centering
    \includegraphics[width=0.95\linewidth]{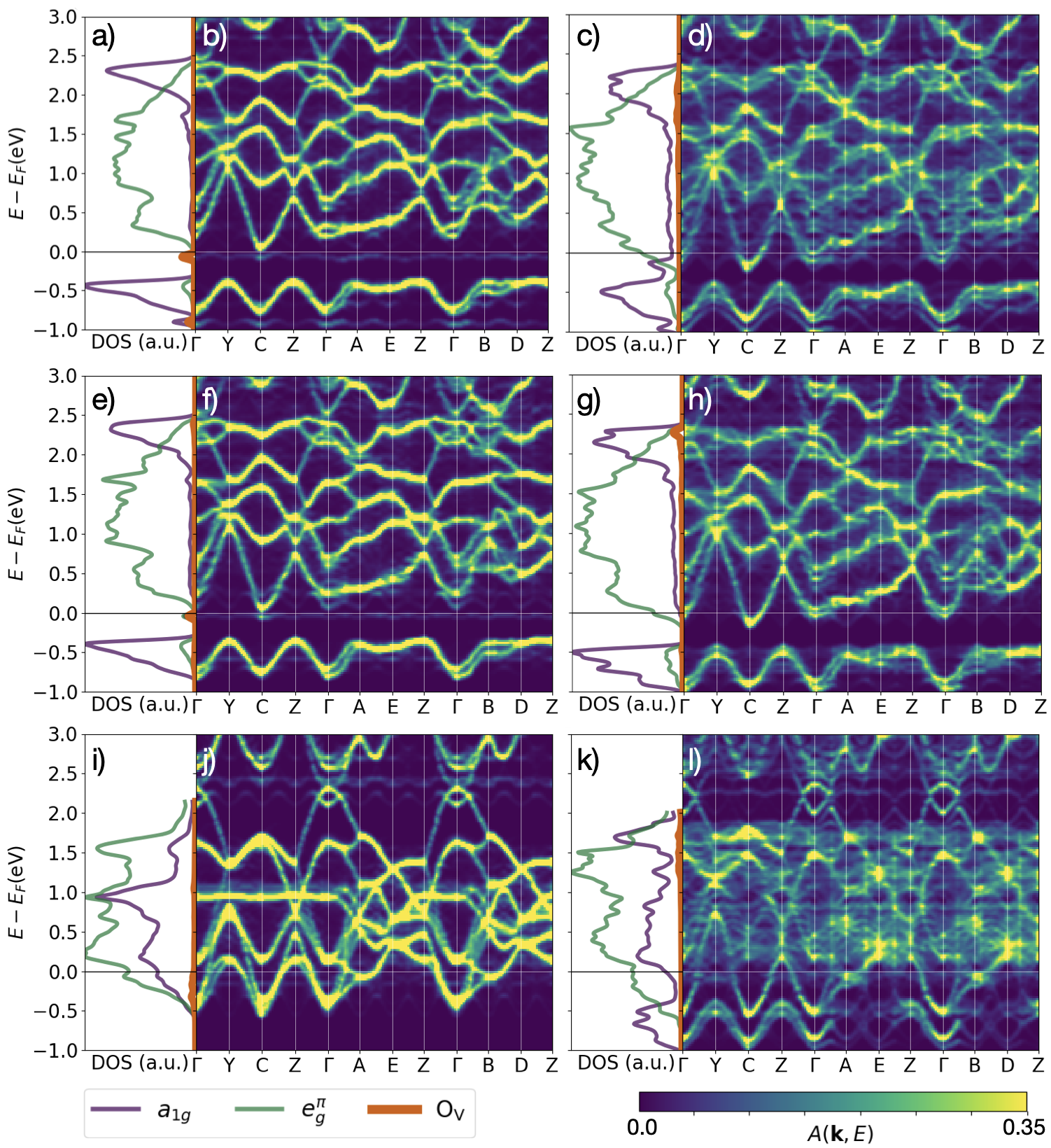}
    \caption{Unfolded band structures and PDOS of the (a-d) M1-O$_\text{SB}$, (e-h) M1-O$_\text{LB}$, and (i-l) R-O configuration before (a, b, e, f, i, j) and (c, d, g, h, k, l) after internal relaxation. Colors in the PDOS denote orbital character: purple ($a_{1g}$), green ($e_g^{\pi}$), and orange (O$_V$). Colors in the band diagram denote spectral weight $A(\mathbf{k},E)$ of the unfolded band structure.}
    \label{fig:BS}
\end{figure*}

In \pref{fig:BS}(a, e), we show the PDOS corresponding to the V $t_{2g}$ subspace and the vacancy WF for the M1-O$_\text{SB}$ and M1-O$_\text{LB}$ configurations, together with the corresponding unfolded band structure in \pref{fig:BS}(b, f), respectively. Similar to the case of the primitive cell, the electronic structures exhibit a strongly localized vacancy state appearing just below the Fermi level, $E_F$ [see orange peak in \pref{fig:BS}(a, e)], but no other major changes compared to stoichiometric VO$_2$ [\pref{fig:pristine}(a, b)]. The unfolded band structure confirms the presence of an energetically isolated intragap state, with a very weakly visible band seen close to the conduction band at the C point in \pref{fig:BS}(b, f). Crucially, the phase remains insulating with a very small gap. The R-O configuration [\pref{fig:BS}(i, j)] remains similar to that of stoichiometric R VO$_2$ [\pref{fig:pristine}(c, d)], remaining metallic, with highly coherent unfolded band structure and a broad vacancy peak just below $E_F$.

\subsection{Relaxed atomic positions}

After internal relaxation, the energy difference between the dimerized and non-dimerized phases is drastically reduced. The high-temperature R phase is stabilized relative to the stoichiometric configuration, from 75.55\,meV/f.u. to 32.27\,meV/f.u. (\pref{tab:energies}, bottom). This finding is consistent with the observed persistence of the R structure to lower temperature. The M1-O$_\text{LB}$ structure remains the most stable, followed closely by M1-O$_\text{SB}$. The small energy difference between the two vacancy configurations makes both configurations plausible in real systems. 

\begin{figure}
    \centering
    \includegraphics[width=\linewidth]{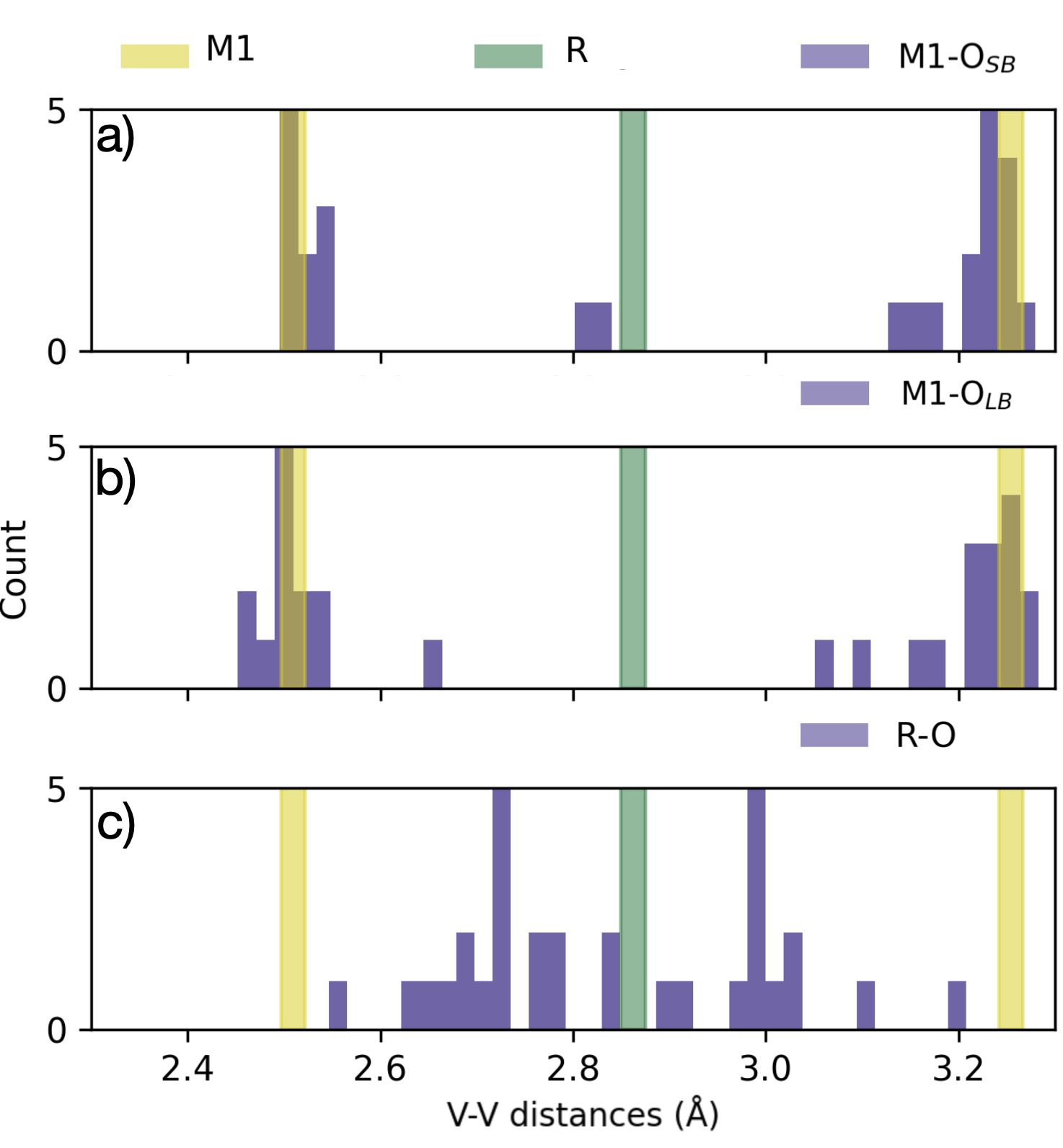}
    \caption{Nearest-neighbor V--V distances (purple) in the internally relaxed (a) M1-O$_\text{SB}$, (b) M1-O$_\text{LB}$, and (c) R-O structures. The stoichiometric reference values for M1 and R are shown in yellow and green, respectively.}
    \label{fig:VV_dists}
\end{figure}

In \pref{fig:VV_dists}(a-c), we show the distribution of nearest-neighbor V--V distances as a measure of the structural changes. In yellow, we show the short and long nearest-neighbor V--V distances of the stoichiometric relaxed M1 structure (at 2.50\,Å and 3.24\,Å), and in green those of the R structure (at 2.85\,Å).

Both monoclinic structures [\pref{fig:VV_dists}(a, b)] display only relatively small and localized structural perturbations. The majority of V--V distances deviate by less than 0.2\,Å from the stoichiometric distances. The M1-O$_\text{LB}$ configuration maintains clear alternation of short and long V--V bonds centered around the stoichiometric values, indicating a persistent dimerized ground state. In M1-O$_\text{SB}$, larger deviations are observed along the oxygen-deficient chain, including compressions and elongations relative to the dimerized reference. Since the dimerized structures have been shown to be the lower energy structures, the small energetic preference for the O$_\text{LB}$ vacancy location could be explained by the energetic preference for the more unperturbed, i.e., dimerized, M1-O$_\text{LB}$ structure. 

In contrast to the M1 configurations, in the R-O configuration [\pref{fig:VV_dists}(c)], V--V distances are perturbed throughout the whole cell, and range broadly from 2.5 to 3.2\,Å. This variation occurs without discernible spatial correlation, indicating strong inter-chain communication of vacancy-induced structural effects in the R phase. A possible explanation of this striking difference compared to the M1 symmetry is the reduced degree of on-chain hybridization (i.e., bonding singlet states), making the R structure more susceptible to interchain perturbations. Notably, the R-O structure does not relax into either of the lower-energy oxygen-deficient structures, indicating this phase to be a local minimum configuration.

Although some V--V distances in M1-O$_\text{SB}$ approach the characteristic R V--V distances [\pref{fig:VV_dists}(a), at around 2.8\,Å], we do not interpret this as a structural transition to the R phase, as has been proposed elsewhere \cite{Chen_et_al:2016b, Zhang_et_al:2017}. First, as we have just seen, the R-O structure does not feature any tendency to dimerize. Secondly, the strongly perturbed, R-like bond lengths present in the M1-O$_\text{SB}$ configuration are in fact localized strictly on the vacancy chain, with no evidence of de-dimerization in adjacent chains.

\begin{figure}
    \centering
    \includegraphics[width=\linewidth]{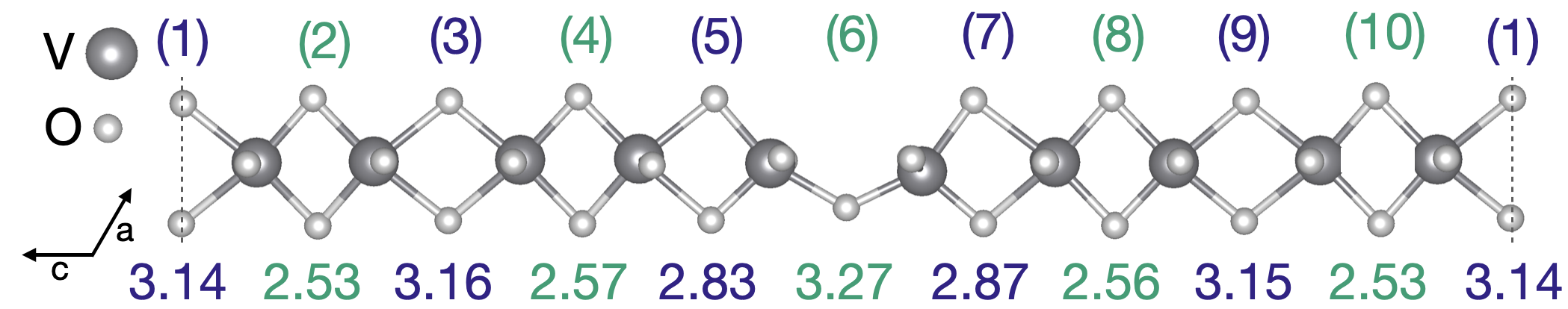}
    \caption{The 10-atom vanadium chain in the relaxed M1-O$_\text{SB}$ structure of the $1\times1\times5$ cell (dashed lines) with bond lengths (in Å) for the initial long (blue) and short (green) V--V bonds. O$_V$ lies on bond (6). V (O) atoms shown in (light) gray.}
    \label{fig:511chain}
\end{figure}

To further rule out global de-dimerization as a result of oxygen deficiency, we additionally consider a $1 \times 1 \times 5$ supercell (VO$_{2-\delta}$ with $\delta=0.05$) containing a periodicity across 10 atoms within the vanadium chains (see \pref{fig:511chain}), as opposed to the 4 atom periodicity along $c$ in the $2\times2\times2$ supercell. We perform an internal relaxation of the M1-O$_\text{SB}$ structure in this supercell and investigate the resultant bond lengths, labeled (1)-(10).

In this structure, we confirm that upon the introduction of a vacancy [at position (6)], we only observe significant changes in bond lengths directly adjacent to the vacancy [(5) and (7)]. In fact, due to the presence of the vacancy, we observe an elongation of the SB hosting the vacancy to 3.27\,Å from the stoichiometric 2.50\,Å, and a compression of the surrounding LB bond lengths to 2.83\,Å and 2.87\,Å from the stoichiometric 3.24\,Å. Dimerization remains intact throughout the rest of the chain. 

These results suggest that the appearance of R-like distances should be interpreted as a localized structural feature induced by the vacancy, rather than an indication of a global structural transition away from the dimerized state. 

We now consider the electronic structure of the relaxed configurations. In \pref{fig:BS}(c, g), we show the PDOS, together with the corresponding unfolded band structure [\pref{fig:BS}(d, h)] of the relevant states around the Fermi level in M1-O$_\text{SB}$ and M1-O$_\text{LB}$, respectively, similar to \pref{fig:BS}(a, b, e, f).

Focusing first on the lowest-energy M1-O$_\text{LB}$ configuration, comparing \pref{fig:BS}(g) to \pref{fig:BS}(e), we see that upon relaxation, the vacancy state (orange) empties and shifts upward to approximately 2.3\,eV above the Fermi level, effectively donating two electrons to the conduction band. Despite the local lattice distortions, the bonding $a_{1g}$ band remains narrow and well-separated from the conduction band [see the $a_{1g}$ peak in \pref{fig:BS}(g)]. The doped system behaves electronically as an almost undisturbed host retaining its original band gap between the bonding $a_{1g}$ and the higher lying bands [\pref{fig:BS}(h)], with metallicity emerging from the conduction band filling resembling a rigid band doping, rather than from defect-induced states or band re-hybridization.

The M1-O$_\text{SB}$ configuration [\pref{fig:BS}(c, d)] shows a similar shift of the vacancy state, with a less rigid response to the distortions, leading to more incoherent weight in the unfolded band structure. The doping electrons populate mostly the $e_g^\pi$ states, even though the bonding $a_{1g}$ states get broadened and smear out the gap, indicating a slight departure from the clean band-filling picture observed in M1-O$_\text{LB}$. 
In contrast, in the R–O configuration [\pref{fig:BS}(k, l)], the more delocalized lattice distortions lead to a broader  $t_{2g}$ manifold and a rather incoherent unfolded band structure. The vacancy state forms a diffuse peak well above $E_F$, while an almost split-off $a_{1g}$ feature appears below. However, no stable electronic motif emerges, and the phase remains high in energy.  

\subsection{Comparison with background charge doping}

\begin{figure}
    \centering
    \includegraphics[width=1\linewidth]{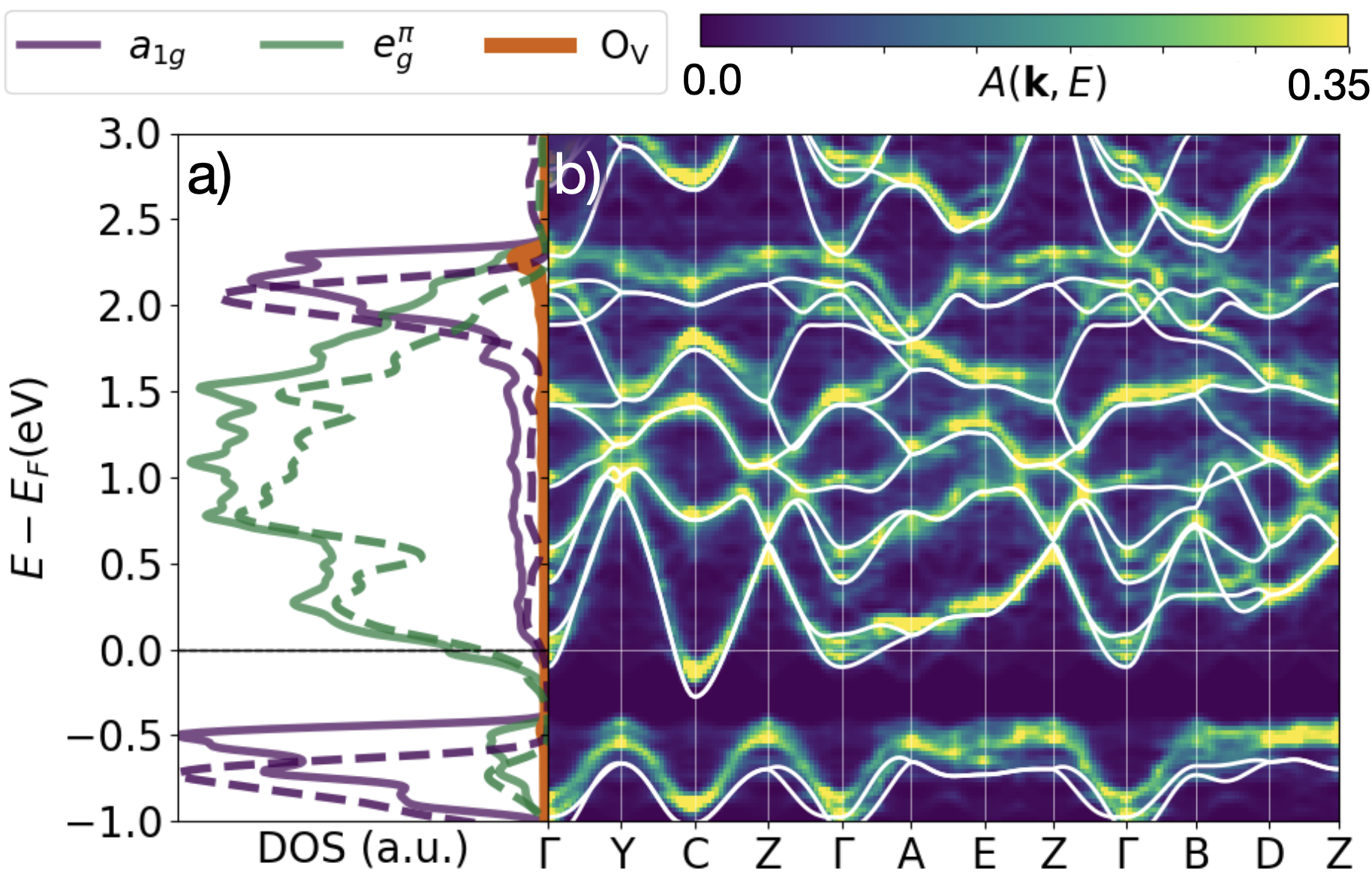}
    \caption{(a) PDOS of the relaxed M1-O$_\text{LB}$ structure with purple (green) lines indicating the $a_{1g}$ ($e_g^\pi$) orbital. O$_V$ shown in orange. Dashed lines correspond to the background charge method with total charge set to $-0.25$. (b) Unfolded band structure of the relaxed M1-O$_\text{LB}$ structure shown in color denoting the spectral weight $A(\mathbf{k},E)$. Band structure corresponding to the background charge method with total charge set to $-0.25$ shown in white lines.}
    \label{fig:edope}
\end{figure}

To further analyze the effects of oxygen vacancies, we now utilize the background charge method, i.e., we consider a stoichiometric primitive M1 unit cell, which we structurally relax with an added electron filling of $-0.25$ elementary charges, corresponding exactly to the number of electrons added by one oxygen vacancy in the $2\times2\times2$ supercell. This extra charge is then compensated in the calculation by adding an equivalent homogeneous positive background charge. In \pref{fig:edope}(a, b), we compare the resulting PDOS and band structure with that of the relaxed, lowest-energy, M1-O$_\text{LB}$ configuration from \pref{fig:BS}(g, h). 

First, we note that, compared to the stoichiometric M1 structure [\pref{fig:pristine}(a, b)], the band structure obtained from the calculation with added electrons essentially corresponds to a rigid shift of the Fermi level upwards in energy, i.e., the behavior of the system under the background charge method strongly resembles a rigid band doping. In particular, we observe no tendency for structural de-dimerization, also reflected in the clearly separated $a_{1g}$ peaks in \pref{fig:edope}(a).

Comparing now to the M1-O$_\text{LB}$ structure, we can again observe a strong resemblance of the corresponding band structures and confirm the simple conduction band filling, as mentioned earlier. Indeed, we observe an excellent agreement between the shifted conduction band due to the oxygen vacancy and the background charge, i.e., a rigid doping (note particularly the energy range between 0 and 2\,eV). These bands specifically correspond to the non-dimerizing $e_g^\pi$ states [green in \pref{fig:edope}(a)]. On the other hand, there is a slight shift of the $a_{1g}$ bands relative to the $e_g^\pi$ conduction bands between the two methods [see the discrepancy around $-0.5$\,eV and 2\,eV in purple in \pref{fig:edope}(a)], while the bonding-antibonding splitting of the $a_{1g}$ states remains very similar.
 
These results directly show that the effect of the oxygen vacancies can, to a good approximation, be interpreted as a rigid doping of the conduction band, formed mainly by the $e_g^\pi$ states, while the overall structure, and in particular the V--V dimerization, remains relatively unperturbed. This is especially true in the M1-O$_\text{LB}$ structure, where the rigid band approximation holds most accurately. 


\section{Summary and Outlook}

In summary, we have investigated the influence of oxygen vacancies on the structural and electronic properties of the rutile R and monoclinic M1 VO$_2$ phases using DFT$+V$. Our results demonstrate that the dimerized M1 structure remains energetically favored in the presence of oxygen vacancies, although the energy penalty for de-dimerization is reduced relative to stoichiometric VO$_2$. This finding aligns with the experimentally observed reduction of the structural transition temperature between the R and M1 phases in VO$_{2-\delta}$ compared to VO$_2$. 

Specifically, our findings suggest a dual mechanism of the MIT suppression: structurally, through a lowered structural phase transition temperature, and electronically, via vacancy-electron doping of the conduction band. Analysis of relaxed oxygen-deficient structures shows that vacancies induce locally constrained distortions without promoting global de-dimerization. Electronically, vacancies introduce defect states below the Fermi level in unrelaxed structures, which become depleted upon structural relaxation and transfer their electrons into the conduction band. This results in a partially filled conduction band and a metallic ground state in the monoclinic phase, regardless of vacancy position, identical to the observed carrier doping effects~\cite{Schrecongost_et_al:2019}. The electronic response agrees very well with a calculation with an added background charge, and closely follows a rigid band doping model of the conduction band $e_g^\pi$ states, highlighting the intrinsic robustness of the M1 structure against vacancy-induced perturbations. Notably, the fact that the dimerization appears unaffected by the doping of the $e_g^\pi$ states indicates agreement with a Peierls picture hosting essentially decoupled electronic degrees of freedom between the dimerizing and non-dimerizing orbitals.

Whether electron delocalization and conduction band injection persist under stronger electron correlation remains an open question, warranting further study either via explicit vacancy-site correlation treatments such as through on-site Hubbard terms applied to vacancy orbitals~\cite{Carta_et_al:2025} or DFT+DMFT techniques~\cite{Souto-Casares/Spaldin/Ederer:2019, Souto-Casares/Spaldin/Ederer:2021}.


\section*{Acknowledgments}
The authors acknowledge useful discussions with A. Carta and I. Timrov. This work was supported by ETH Z\"{u}rich and the Swiss National Science Foundation (Grant No.~209454). Calculations were performed on the ETH Z\"{u}rich Euler cluster and the Swiss National Supercomputing Center Eiger cluster under Project ID s1304.

\appendix

\section{Stoichiometric R and M1 structures}

\begin{figure}
    \centering
    \includegraphics[width=0.85\linewidth]{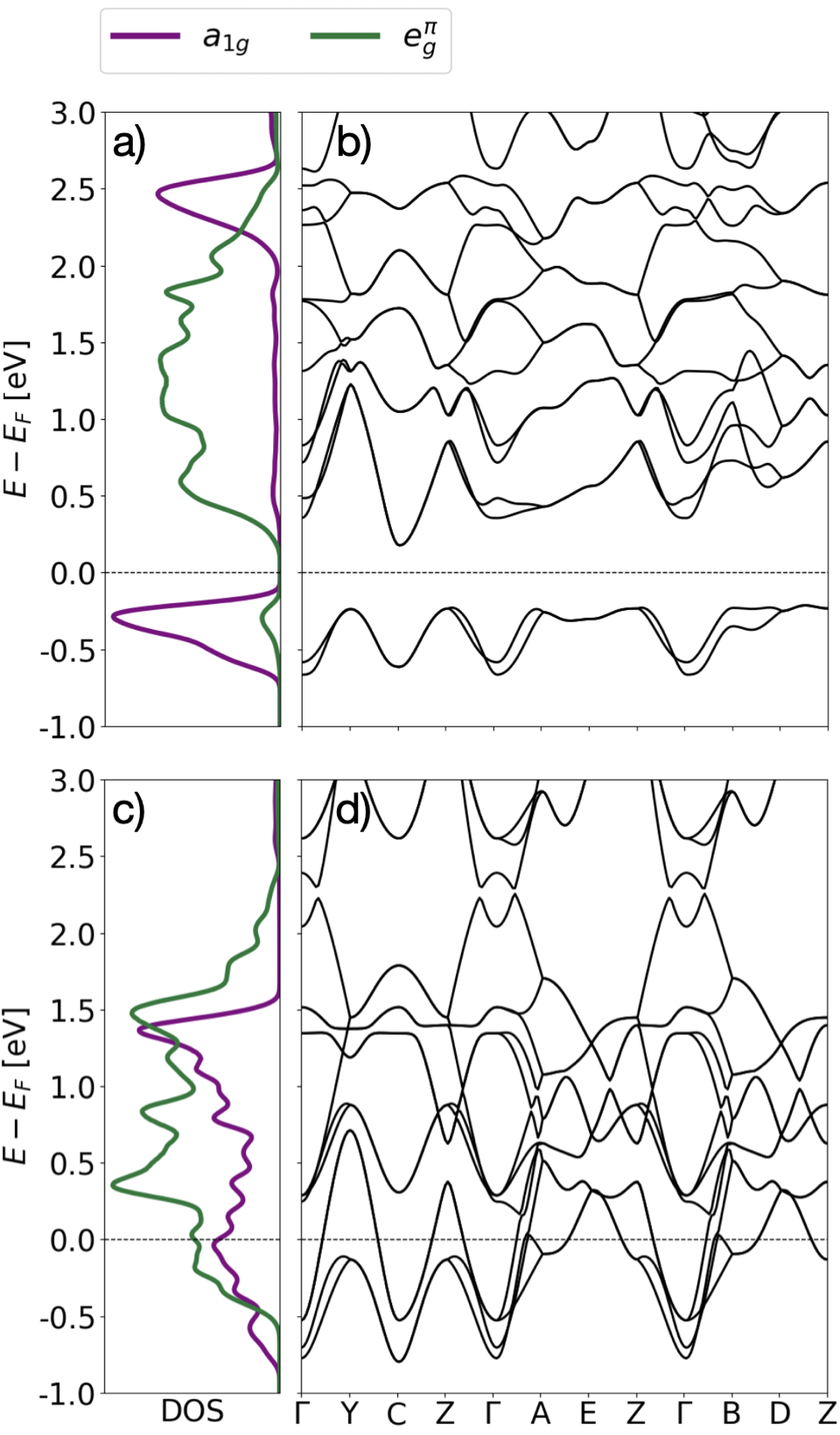}
    \caption{PDOS and band structures for the stoichiometric M1 (a, b) and R (c, d) internally relaxed structures using $V$=1, with $a_{1g}$ and $e_g^\pi$ PDOS in purple and green respectively.}
    \label{fig:pristine}
\end{figure}

In \pref{fig:pristine} we show the PDOS and band structures for the stoichiometric M1 and R structures.

\newpage

\bibliography{ovac}

@article{Aetukuri_et_al:2013,
  title = {Control of the Metal--Insulator Transition in Vanadium Dioxide by Modifying Orbital Occupancy},
  author = {Aetukuri, Nagaphani B. and Gray, Alexander X. and Drouard, Marc and Cossale, Matteo and Gao, Li and Reid, Alexander H. and Kukreja, Roopali and Ohldag, Hendrik and Jenkins, Catherine A. and Arenholz, Elke and Roche, Kevin P. and D{\"u}rr, Hermann A. and Samant, Mahesh G. and Parkin, Stuart S. P.},
  year = {2013},
  month = oct,
  journal = {Nat. Phys.},
  volume = {9},
  number = {10},
  pages = {661--666},
  publisher = {Nature Publishing Group},
  issn = {1745-2481},
  doi = {10.1038/nphys2733},
  urldate = {2025-04-09},
  copyright = {2013 Springer Nature Limited}
}

@article{Appavoo_et_al:2012,
  title = {Role of {{Defects}} in the {{Phase Transition}} of {{VO}}{\textsubscript{2}} {{Nanoparticles Probed}} by {{Plasmon Resonance Spectroscopy}}},
  author = {Appavoo, Kannatassen and Lei, Dang Yuan and Sonnefraud, Yannick and Wang, Bin and Pantelides, Sokrates T. and Maier, Stefan A. and Haglund, Richard F. Jr.},
  year = {2012},
  month = feb,
  journal = {Nano Lett.},
  volume = {12},
  number = {2},
  pages = {780--786},
  publisher = {American Chemical Society},
  issn = {1530-6984},
  doi = {10.1021/nl203782y},
  urldate = {2025-08-10}
}

@article{Atkin_et_al:2012,
  title = {Strain and Temperature Dependence of the Insulating Phases of {{VO}}{\textsubscript{2}}  near the Metal-Insulator Transition},
  author = {Atkin, Joanna M. and Berweger, Samuel and Chavez, Emily K. and Raschke, Markus B. and Cao, Jinbo and Fan, Wen and Wu, Junqiao},
  year = {2012},
  month = jan,
  journal = {Phys. Rev. B},
  volume = {85},
  number = {2},
  pages = {020101},
  issn = {1098-0121, 1550-235X},
  doi = {10.1103/PhysRevB.85.020101},
  urldate = {2024-01-08}
}

@article{Basyooni_et_al:2022,
  title = {Tuning the {{Metal}}--{{Insulator Transition Properties}} of {{VO}}{\textsubscript{2}} {{Thin Films}} with the {{Synergetic Combination}} of {{Oxygen Vacancies}}, {{Strain Engineering}}, and {{Tungsten Doping}}},
  author = {Basyooni, Mohamed A. and {Al-Dossari}, Mawaheb and Zaki, Shrouk E. and Eker, Yasin Ramazan and Yilmaz, Mucahit and Shaban, Mohamed},
  year = {2022},
  month = jan,
  journal = {Nanomater.},
  volume = {12},
  number = {9},
  pages = {1470},
  publisher = {Multidisciplinary Digital Publishing Institute},
  issn = {2079-4991},
  doi = {10.3390/nano12091470},
  urldate = {2025-08-10},
  copyright = {http://creativecommons.org/licenses/by/3.0/}
}

@article{Bianconi:1982,
  title = {Multiplet Splitting of Final-State Configurations in x-Ray-Absorption Spectrum of Metal {{VO}}{\textsubscript{2}}: {{Effect}} of Core-Hole-Screening, Electron Correlation, and Metal-Insulator Transition},
  author = {Bianconi, Antonio},
  year = {1982},
  month = sep,
  journal = {Phys. Rev. B},
  volume = {26},
  number = {6},
  pages = {2741--2747},
  publisher = {American Physical Society},
  doi = {10.1103/PhysRevB.26.2741},
  urldate = {2025-08-10}
}

@article{Biermann_et_al:2005,
  title = {Dynamical Singlets and Correlation-Assisted {{Peierls}} Transition in {{VO}}{\textsubscript{2}}},
  author = {Biermann, S. and Poteryaev, A. and Lichtenstein, A. I. and Georges, A.},
  year = {2005},
  month = jan,
  journal = {Phys. Rev. Lett.},
  volume = {94},
  number = {2},
  pages = {026404},
  issn = {0031-9007, 1079-7114},
  doi = {10.1103/PhysRevLett.94.026404},
  urldate = {2022-02-17}
}

@article{Boykin/Klimeck:2005,
  title = {Practical Application of Zone-Folding Concepts in Tight-Binding Calculations},
  author = {Boykin, Timothy B. and Klimeck, Gerhard},
  year = {2005},
  month = mar,
  journal = {Phys. Rev. B},
  volume = {71},
  number = {11},
  pages = {115215},
  issn = {1098-0121, 1550-235X},
  doi = {10.1103/PhysRevB.71.115215},
  urldate = {2022-07-20}
}

@article{Brito_et_al:2016,
  title = {Metal-Insulator Transition in {{VO}}{\textsubscript{2}}: {{A DFT}}+{{DMFT}} Perspective},
  author = {Brito, W. H. and Aguiar, M. C. O. and Haule, K. and Kotliar, G.},
  year = {2016},
  month = jul,
  journal = {Phys. Rev. Lett.},
  volume = {117},
  number = {5},
  pages = {056402},
  issn = {0031-9007, 1079-7114},
  doi = {10.1103/PhysRevLett.117.056402},
  urldate = {2022-02-04}
}

@article{Cao_et_al:2010,
  title = {Extended Mapping and Exploration of the Vanadium Dioxide Stress-Temperature Phase Diagram},
  author = {Cao, J. and Gu, Y. and Fan, W. and Chen, L. Q. and Ogletree, D. F. and Chen, K. and Tamura, N. and Kunz, M. and Barrett, C. and Seidel, J. and Wu, J.},
  year = {2010},
  month = jul,
  journal = {Nano Lett.},
  volume = {10},
  number = {7},
  pages = {2667--2673},
  publisher = {American Chemical Society},
  issn = {1530-6984},
  doi = {10.1021/nl101457k},
  urldate = {2024-11-05}
}

@article{Carta_et_al:2025,
  title = {Explicit Demonstration of the Equivalence between {{DFT}}+{{{\emph{U}}}} and the {{Hartree-Fock}} Limit of {{DFT}}+{{DMFT}}},
  author = {Carta, Alberto and Timrov, Iurii and Mlkvik, Peter and Hampel, Alexander and Ederer, Claude},
  year = {2025},
  month = mar,
  journal = {Phys. Rev. Res.},
  volume = {7},
  number = {1},
  pages = {013289},
  publisher = {American Physical Society},
  doi = {10.1103/PhysRevResearch.7.013289},
  urldate = {2025-03-19}
}

@article{Chen_et_al:2016a,
  title = {Tuning the Phase Transition Temperature, Electrical and Optical Properties of {{VO}}{\textsubscript{2}} by Oxygen Nonstoichiometry: Insights from First-Principles Calculations},
  author = {Chen, Lanli and Wang, Xiaofang and Wan, Dongyun and Cui, Yuanyuan and Liu, Bin and Shi, Siqi and Luo, Hongjie and Gao, Yanfeng},
  year = {2016},
  journal = {RSC Adv.},
  volume = {6},
  number = {77},
  pages = {73070--73082},
  issn = {2046-2069},
  doi = {10.1039/C6RA09449J},
  urldate = {2022-02-04}
}

@article{Chen_et_al:2016b,
  title = {The {{Dynamic Phase Transition Modulation}} of {{Ion-Liquid Gating VO}}{\textsubscript{2}} {{Thin Film}}: {{Formation}}, {{Diffusion}}, and {{Recovery}} of {{Oxygen Vacancies}}},
  author = {Chen, Shi and jun Wang, Xi and Fan, Lele and Liao, Guangming and Chen, Yuliang and Chu, Wangsheng and Song, Li and Jiang, Jun and Zou, Chongwen},
  year = {2016},
  journal = {Adv. Funct. Mater},
  volume = {26},
  number = {20},
  pages = {3532--3541},
  issn = {1616-3028},
  doi = {10.1002/adfm.201505399},
  urldate = {2025-08-10},
  copyright = {{\copyright} 2016 WILEY-VCH Verlag GmbH \& Co. KGaA, Weinheim}
}

@article{Cui_et_al:2018,
  title = {Thermochromic {{VO}}{\textsubscript{2}} for Energy-Efficient Smart Windows},
  author = {Cui, Yuanyuan and Ke, Yujie and Liu, Chang and Chen, Zhang and Wang, Ning and Zhang, Liangmiao and Zhou, Yang and Wang, Shancheng and Gao, Yanfeng and Long, Yi},
  year = {2018},
  month = sep,
  journal = {Joule},
  volume = {2},
  number = {9},
  pages = {1707--1746},
  issn = {2542-4351},
  doi = {10.1016/j.joule.2018.06.018},
  urldate = {2022-05-27}
}

@article{Eyert:2002a,
  title = {The Metal-Insulator Transitions of {{VO}}{\textsubscript{2}}: {{A}} Band Theoretical Approach},
  author = {Eyert, V.},
  year = {2002},
  month = oct,
  journal = {Ann. Phys.},
  volume = {11},
  number = {9},
  pages = {650--704},
  issn = {00033804, 15213889},
  doi = {10.1002/1521-3889(200210)11:9<650::AID-ANDP650>3.0.CO;2-K},
  urldate = {2022-02-09}
}

@article{Fan_et_al:2013a,
  title = {Growth and Phase Transition Characteristics of Pure {{M-phase VO}}{\textsubscript{2}} Epitaxial Film Prepared by Oxide Molecular Beam Epitaxy},
  author = {Fan, L. L. and Chen, S. and Wu, Y. F. and Chen, F. H. and Chu, W. S. and Chen, X. and Zou, C. W. and Wu, Z. Y.},
  year = {2013},
  month = sep,
  journal = {Appl. Phys. Lett.},
  volume = {103},
  number = {13},
  pages = {131914},
  issn = {0003-6951},
  doi = {10.1063/1.4823511},
  urldate = {2025-10-23}
}

@article{Fan_et_al:2018,
  title = {Revealing the Role of Oxygen Vacancies on the Phase Transition of {{VO}}{\textsubscript{2}} Film from the Optical-Constant Measurements},
  author = {Fan, Lele L. and Wang, Xiangqi and Wang, Feng and Zhang, Qinfang and Zhu, Lei and Meng, Qiangqiang and Wang, Baolin and Zhang, Zengming and Zou, Chongwen},
  year = {2018},
  month = may,
  journal = {RSC Adv.},
  volume = {8},
  number = {34},
  pages = {19151--19156},
  publisher = {The Royal Society of Chemistry},
  issn = {2046-2069},
  doi = {10.1039/C8RA03292K},
  urldate = {2025-04-09}
}

@article{Ganesh_et_al:2020,
  title = {Doping a Bad Metal: {{Origin}} of Suppression of the Metal-Insulator Transition in Nonstoichiometric {{VO}}{\textsubscript{2}}},
  author = {Ganesh, P. and Lechermann, Frank and Kyl{\"a}np{\"a}{\"a}, Ilkka and Krogel, Jaron T. and Kent, Paul R. C. and Heinonen, Olle},
  year = {2020},
  month = apr,
  journal = {Phys. Rev. B},
  volume = {101},
  number = {15},
  pages = {155129},
  issn = {2469-9950, 2469-9969},
  doi = {10.1103/PhysRevB.101.155129},
  urldate = {2022-03-07}
}

@article{Garrity_et_al:2014,
  title = {Pseudopotentials for High-Throughput {{DFT}} Calculations},
  author = {Garrity, Kevin F. and Bennett, Joseph W. and Rabe, Karin M. and Vanderbilt, David},
  year = {2014},
  month = jan,
  journal = {Comput. Mater. Sci.},
  volume = {81},
  pages = {446--452},
  issn = {0927-0256},
  doi = {10.1016/j.commatsci.2013.08.053},
  urldate = {2023-07-17}
}

@article{Gatti_et_al:2007,
  title = {Understanding Correlations in Vanadium Dioxide from First Principles},
  author = {Gatti, Matteo and Bruneval, Fabien and Olevano, Valerio and Reining, Lucia},
  year = {2007},
  month = dec,
  journal = {Phys. Rev. Lett.},
  volume = {99},
  number = {26},
  pages = {266402},
  issn = {0031-9007, 1079-7114},
  doi = {10.1103/PhysRevLett.99.266402},
  urldate = {2022-05-09}
}

@article{Giannozzi_et_al:2009,
  title = {{{QUANTUM ESPRESSO}}: A Modular and Open-Source Software Project for Quantum Simulations of Materials},
  author = {Giannozzi, Paolo and Baroni, Stefano and Bonini, Nicola and Calandra, Matteo and Car, Roberto and Cavazzoni, Carlo and Ceresoli, Davide and Chiarotti, Guido L. and Cococcioni, Matteo and Dabo, Ismaila and Corso, Andrea Dal and de Gironcoli, Stefano and Fabris, Stefano and Fratesi, Guido and Gebauer, Ralph and Gerstmann, Uwe and Gougoussis, Christos and Kokalj, Anton and Lazzeri, Michele and {Martin-Samos}, Layla and Marzari, Nicola and Mauri, Francesco and Mazzarello, Riccardo and Paolini, Stefano and Pasquarello, Alfredo and Paulatto, Lorenzo and Sbraccia, Carlo and Scandolo, Sandro and Sclauzero, Gabriele and Seitsonen, Ari P. and Smogunov, Alexander and Umari, Paolo and Wentzcovitch, Renata M.},
  year = {2009},
  month = sep,
  journal = {J. Phys.: Condens. Matter},
  volume = {21},
  number = {39},
  pages = {395502},
  issn = {0953-8984},
  doi = {10.1088/0953-8984/21/39/395502},
  urldate = {2023-07-17}
}

@article{Giannozzi_et_al:2017,
  title = {Advanced Capabilities for Materials Modelling with {{Quantum ESPRESSO}}},
  author = {Giannozzi, P and Andreussi, O and Brumme, T and Bunau, O and Buongiorno Nardelli, M and Calandra, M and Car, R and Cavazzoni, C and Ceresoli, D and Cococcioni, M and Colonna, N and Carnimeo, I and Dal Corso, A and {de Gironcoli}, S and Delugas, P and DiStasio, R A and Ferretti, A and Floris, A and Fratesi, G and Fugallo, G and Gebauer, R and Gerstmann, U and Giustino, F and Gorni, T and Jia, J and Kawamura, M and Ko, H-Y and Kokalj, A and K{\"u}{\c c}{\"u}kbenli, E and Lazzeri, M and Marsili, M and Marzari, N and Mauri, F and Nguyen, N L and Nguyen, H-V and {Otero-de-la-Roza}, A and Paulatto, L and Ponc{\'e}, S and Rocca, D and Sabatini, R and Santra, B and Schlipf, M and Seitsonen, A P and Smogunov, A and Timrov, I and Thonhauser, T and Umari, P and Vast, N and Wu, X and Baroni, S},
  year = {2017},
  month = nov,
  journal = {J. Phys.: Condens. Matter},
  volume = {29},
  number = {46},
  pages = {465901},
  issn = {0953-8984, 1361-648X},
  doi = {10.1088/1361-648X/aa8f79},
  urldate = {2022-05-09}
}

@article{Giannozzi_et_al:2020,
  title = {Quantum {{ESPRESSO}} toward the Exascale},
  author = {Giannozzi, Paolo and Baseggio, Oscar and Bonf{\`a}, Pietro and Brunato, Davide and Car, Roberto and Carnimeo, Ivan and Cavazzoni, Carlo and {de Gironcoli}, Stefano and Delugas, Pietro and Ferrari Ruffino, Fabrizio and Ferretti, Andrea and Marzari, Nicola and Timrov, Iurii and Urru, Andrea and Baroni, Stefano},
  year = {2020},
  month = apr,
  journal = {J. Chem. Phys.},
  volume = {152},
  number = {15},
  pages = {154105},
  issn = {0021-9606},
  doi = {10.1063/5.0005082},
  urldate = {2025-08-10}
}

@article{Goodenough:1971,
  title = {The Two Components of the Crystallographic Transition in {{VO}}{\textsubscript{2}}},
  author = {Goodenough, John B.},
  year = {1971},
  month = nov,
  journal = {J. Solid State Chem.},
  volume = {3},
  number = {4},
  pages = {490--500},
  issn = {00224596},
  doi = {10.1016/0022-4596(71)90091-0},
  urldate = {2022-02-17}
}

@article{Grandi/Amaricci/Fabrizio:2020,
  title = {Unraveling the {{Mott-Peierls}} Intrigue in Vanadium Dioxide},
  author = {Grandi, F. and Amaricci, A. and Fabrizio, M.},
  year = {2020},
  month = mar,
  journal = {Phys. Rev. Res.},
  volume = {2},
  number = {1},
  pages = {013298},
  issn = {2643-1564},
  doi = {10.1103/PhysRevResearch.2.013298},
  urldate = {2022-02-17}
}

@article{Guo_et_al:2024,
  title = {Effects of Oxygen Vacancies and Interfacial Strain on the Metal--Insulator Transition of {{VO}}{\textsubscript{2}} Nanobeams},
  author = {Guo, Xitao and Liu, Xin and Zafar, Zainab and Cheng, Guiquan and Li, Yunhai and Nan, Haiyan and Lin, Lianghua and Zou, Jijun},
  year = {2024},
  month = apr,
  journal = {Phys. Chem. Chem. Phys.},
  volume = {26},
  number = {14},
  pages = {10737--10745},
  publisher = {The Royal Society of Chemistry},
  issn = {1463-9084},
  doi = {10.1039/D3CP06040C},
  urldate = {2025-08-10}
}

@article{Haas_et_al:2024,
  title = {Incorporating Static Intersite Correlation Effects in Vanadium Dioxide through {{DFT}}+{{{\emph{V}}}}},
  author = {Haas, Lea and Mlkvik, Peter and Spaldin, Nicola A. and Ederer, Claude},
  year = {2024},
  month = nov,
  journal = {Phys. Rev. Res.},
  volume = {6},
  number = {4},
  pages = {043177},
  publisher = {American Physical Society},
  doi = {10.1103/PhysRevResearch.6.043177},
  urldate = {2024-12-11}
}

@article{Iraola_et_al:2022,
  title = {{{IrRep}}: {{Symmetry}} Eigenvalues and Irreducible Representations of Ab Initio Band Structures},
  author = {Iraola, Mikel and Ma{\~n}es, Juan L. and Bradlyn, Barry and Horton, Matthew K. and Neupert, Titus and Vergniory, Maia G. and Tsirkin, Stepan S.},
  year = {2022},
  month = mar,
  journal = {Comput. Phys. Commun.},
  volume = {272},
  pages = {108226},
  issn = {0010-4655},
  doi = {10.1016/j.cpc.2021.108226},
  urldate = {2022-05-27}
}

@article{Jeong_et_al:2013,
  title = {Suppression of {{Metal-Insulator Transition}} in {{VO}}{\textsubscript{2}} by {{Electric Field}}--{{Induced Oxygen Vacancy Formation}}},
  author = {Jeong, Jaewoo and Aetukuri, Nagaphani and Graf, Tanja and Schladt, Thomas D. and Samant, Mahesh G. and Parkin, Stuart S. P.},
  year = {2013},
  month = mar,
  journal = {Science},
  volume = {339},
  number = {6126},
  pages = {1402--1405},
  publisher = {American Association for the Advancement of Science},
  doi = {10.1126/science.1230512},
  urldate = {2025-08-10}
}

@article{Kim_et_al:2014,
  title = {Optimization of the Semiconductor-Metal Transition in {{VO}}{\textsubscript{2}} Epitaxial Thin Films as a Function of Oxygen Growth Pressure},
  author = {Kim, H. and Charipar, N. and Osofsky, M. and Qadri, S. B. and Piqu{\'e}, A.},
  year = {2014},
  month = feb,
  journal = {Appl. Phys. Lett.},
  volume = {104},
  number = {8},
  pages = {081913},
  issn = {0003-6951},
  doi = {10.1063/1.4866806},
  urldate = {2025-08-10}
}

@article{Krammer_et_al:2017,
  title = {Elevated Transition Temperature in {{Ge}} Doped {{VO}}{\textsubscript{2}} Thin Films},
  author = {Krammer, Anna and Magrez, Arnaud and Vitale, Wolfgang A. and Mocny, Piotr and Jeanneret, Patrick and Guibert, Edouard and Whitlow, Harry J. and Ionescu, Adrian M. and Sch{\"u}ler, Andreas},
  year = {2017},
  month = jul,
  journal = {J. Appl. Phys.},
  volume = {122},
  number = {4},
  pages = {045304},
  issn = {0021-8979, 1089-7550},
  doi = {10.1063/1.4995965},
  urldate = {2022-02-17}
}

@article{Lee_et_al:2018,
  title = {Isostructural Metal-Insulator Transition in {{VO}}{\textsubscript{2}}},
  author = {Lee, D. and Chung, B. and Shi, Y. and Kim, G.-Y. and Campbell, N. and Xue, F. and Song, K. and Choi, S.-Y. and Podkaminer, J. P. and Kim, T. H. and Ryan, P. J. and Kim, J.-W. and Paudel, T. R. and Kang, J.-H. and Spinuzzi, J. W. and Tenne, D. A. and Tsymbal, E. Y. and Rzchowski, M. S. and Chen, L. Q. and Lee, J. and Eom, C. B.},
  year = {2018},
  month = nov,
  journal = {Science},
  volume = {362},
  number = {6418},
  pages = {1037--1040},
  issn = {0036-8075, 1095-9203},
  doi = {10.1126/science.aam9189},
  urldate = {2022-02-17}
}

@article{LeiriaCampoJr/Cococcioni:2010,
  title = {Extended {{DFT}}+{{{\emph{U}}}}+{{{\emph{V}}}} Method with On-Site and Inter-Site Electronic Interactions},
  author = {Leiria Campo Jr, Vivaldo and Cococcioni, Matteo},
  year = {2010},
  month = feb,
  journal = {J. Phys.: Condens. Matter},
  volume = {22},
  number = {5},
  pages = {055602},
  issn = {0953-8984, 1361-648X},
  doi = {10.1088/0953-8984/22/5/055602},
  urldate = {2022-05-02}
}

@article{Li_et_al:2017,
  title = {Imaging Metal-like Monoclinic Phase Stabilized by Surface Coordination Effect in Vanadium Dioxide Nanobeam},
  author = {Li, Zejun and Wu, Jiajing and Hu, Zhenpeng and Lin, Yue and Chen, Qi and Guo, Yuqiao and Liu, Yuhua and Zhao, Yingcheng and Peng, Jing and Chu, Wangsheng and Wu, Changzheng and Xie, Yi},
  year = {2017},
  month = jun,
  journal = {Nat. Commun.},
  volume = {8},
  number = {1},
  pages = {15561},
  publisher = {Nature Publishing Group},
  issn = {2041-1723},
  doi = {10.1038/ncomms15561},
  urldate = {2025-08-10},
  copyright = {2017 The Author(s)}
}

@article{Liang_et_al:2016,
  title = {Effect of Oxygen Vacancies on the Electronic and Optical Properties of Vanadium Dioxide},
  author = {Liang, R. and Wu, J. and Li, N. and Liu, X.},
  year = {2016},
  month = apr,
  journal = {Optoelectron. Adv. Mat.},
  volume = {10},
  number = {March-April 2016},
  pages = {191--195},
  publisher = {OAM-RC},
  issn = {1842-6573},
  urldate = {2025-08-10}
}

@article{Lu_et_al:2020a,
  title = {Metal--Insulator Transition Tuned by Oxygen Vacancy Migration across {{TiO}}{\textsubscript{2}}/{{VO}}{\textsubscript{2}} Interface},
  author = {Lu, Qiyang and Sohn, Changhee and Hu, Guoxiang and Gao, Xiang and Chisholm, Matthew F. and Kyl{\"a}np{\"a}{\"a}, Ilkka and Krogel, Jaron T. and Kent, Paul R. C. and Heinonen, Olle and Ganesh, P. and Lee, Ho Nyung},
  year = {2020},
  month = oct,
  journal = {Sci. Rep.},
  volume = {10},
  number = {1},
  pages = {18554},
  issn = {2045-2322},
  doi = {10.1038/s41598-020-75695-1},
  urldate = {2024-08-22}
}

@article{Makarevich_et_al:2021,
  title = {Delicate Tuning of Epitaxial {{VO}}{\textsubscript{2}} Films for Ultra-Sharp Electrical and Intense {{IR}} Optical Switching Properties},
  author = {Makarevich, Artem M. and Sobol, Alexander G. and Sadykov, Ilia I. and Sharovarov, Dmitrii I. and Amelichev, Vadim A. and Tsymbarenko, Dmitry M. and Boytsova, Olga V. and Kaul, Andrey R.},
  year = {2021},
  month = feb,
  journal = {J. Alloys Compd.},
  volume = {853},
  pages = {157214},
  issn = {0925-8388},
  doi = {10.1016/j.jallcom.2020.157214},
  urldate = {2025-08-10}
}

@article{Marzari/Vanderbilt:1997,
  title = {Maximally Localized Generalized {{Wannier}} Functions for Composite Energy Bands},
  author = {Marzari, Nicola and Vanderbilt, David},
  year = {1997},
  month = nov,
  journal = {Phys. Rev. B},
  volume = {56},
  number = {20},
  pages = {12847--12865},
  publisher = {American Physical Society},
  doi = {10.1103/PhysRevB.56.12847},
  urldate = {2023-04-20}
}

@article{Matsuda_et_al:2025,
  title = {Microstructural Characterization of Oxygen-Defective {{VO}}{\textsubscript{2-x}} Films Produced by Precisely Controlled Oxidation of Vanadium Metal Foil},
  author = {Matsuda, Mitsuhiro and Nagata, Miho and Akamine, Hiroshi and Shida, Kenji and Matsuda, Motohide},
  year = {2025},
  journal = {J. Ceram. Soc. Jpn.},
  volume = {133},
  number = {1},
  pages = {9--14},
  doi = {10.2109/jcersj2.24088}
}

@article{McWhan_et_al:1974,
  title = {X-Ray Diffraction Study of Metallic {{VO}}{\textsubscript{2}}},
  author = {McWhan, D. B. and Marezio, M. and Remeika, J. P. and Dernier, P. D.},
  year = {1974},
  month = jul,
  journal = {Phys. Rev. B},
  volume = {10},
  number = {2},
  pages = {490--495},
  publisher = {American Physical Society},
  doi = {10.1103/PhysRevB.10.490},
  urldate = {2023-09-11}
}

@article{Medeiros_et_al:2015,
  title = {Unfolding Spinor Wave Functions and Expectation Values of General Operators: {{Introducing}} the Unfolding-Density Operator},
  author = {Medeiros, Paulo V. C. and Tsirkin, Stepan S. and Stafstr{\"o}m, Sven and Bj{\"o}rk, Jonas},
  year = {2015},
  month = jan,
  journal = {Phys. Rev. B},
  volume = {91},
  number = {4},
  pages = {041116},
  publisher = {American Physical Society},
  doi = {10.1103/PhysRevB.91.041116},
  urldate = {2022-02-24}
}

@article{Medeiros/Stafstrom/Bjork:2014,
  title = {Effects of Extrinsic and Intrinsic Perturbations on the Electronic Structure of Graphene: {{Retaining}} an Effective Primitive Cell Band Structure by Band Unfolding},
  author = {Medeiros, Paulo V. C. and Stafstr{\"o}m, Sven and Bj{\"o}rk, Jonas},
  year = {2014},
  month = jan,
  journal = {Phys. Rev. B},
  volume = {89},
  number = {4},
  pages = {041407},
  publisher = {American Physical Society},
  doi = {10.1103/PhysRevB.89.041407},
  urldate = {2022-02-24}
}

@article{Mlkvik_et_al:2024,
  title = {Single-Site {{DFT}}+{{DMFT}} for Vanadium Dioxide Using Bond-Centered Orbitals},
  author = {Mlkvik, Peter and Merkel, Maximilian E. and Spaldin, Nicola A. and Ederer, Claude},
  year = {2024},
  month = jul,
  journal = {Phys. Rev. Res.},
  volume = {6},
  number = {3},
  pages = {033122},
  publisher = {American Physical Society},
  doi = {10.1103/PhysRevResearch.6.033122},
  urldate = {2024-08-19}
}

@article{Mlkvik/Ederer/Spaldin:2022,
  title = {Influence of Germanium Substitution on the Structural and Electronic Stability of the Competing Vanadium Dioxide Phases},
  author = {Mlkvik, Peter and Ederer, Claude and Spaldin, Nicola A.},
  year = {2022},
  month = nov,
  journal = {Phys. Rev. Res.},
  volume = {4},
  number = {4},
  pages = {043129},
  publisher = {American Physical Society},
  doi = {10.1103/PhysRevResearch.4.043129},
  urldate = {2023-07-17}
}

@article{Morin:1959,
  title = {Oxides Which Show a Metal-to-Insulator Transition at the {{N{\'e}el}} Temperature},
  author = {Morin, F. J.},
  year = {1959},
  month = jul,
  journal = {Phys. Rev. Lett.},
  volume = {3},
  number = {1},
  pages = {34--36},
  publisher = {American Physical Society},
  doi = {10.1103/PhysRevLett.3.34},
  urldate = {2025-04-10}
}

@article{Morrison_et_al:2014,
  title = {A Photoinduced Metal-like Phase of Monoclinic {{VO}}{\textsubscript{2}} Revealed by Ultrafast Electron Diffraction},
  author = {Morrison, Vance R. and Chatelain, {\relax Robert}. P. and Tiwari, Kunal L. and Hendaoui, Ali and Bruh{\'a}cs, Andrew and Chaker, Mohamed and Siwick, Bradley J.},
  year = {2014},
  month = oct,
  journal = {Science},
  volume = {346},
  number = {6208},
  pages = {445--448},
  publisher = {American Association for the Advancement of Science},
  doi = {10.1126/science.1253779},
  urldate = {2025-08-10}
}

@article{Najera_et_al:2017,
  title = {Resolving the {{VO}}{\textsubscript{2}} Controversy: {{Mott}} Mechanism Dominates the Insulator-to-Metal Transition},
  author = {N{\'a}jera, O. and Civelli, M. and Dobrosavljevi{\'c}, V. and Rozenberg, M. J.},
  year = {2017},
  month = jan,
  journal = {Phys. Rev. B},
  volume = {95},
  number = {3},
  pages = {035113},
  issn = {2469-9950, 2469-9969},
  doi = {10.1103/PhysRevB.95.035113},
  urldate = {2022-02-04}
}

@article{Passarello_et_al:2016,
  title = {Metallization of {{Epitaxial VO}}{\textsubscript{2}} {{Films}} by {{Ionic Liquid Gating}} through {{Initially Insulating TiO2 Layers}}},
  author = {Passarello, Donata and Altendorf, Simone G. and Jeong, Jaewoo and Samant, Mahesh G. and Parkin, Stuart S. P.},
  year = {2016},
  month = sep,
  journal = {Nano Lett.},
  volume = {16},
  number = {9},
  pages = {5475--5481},
  publisher = {American Chemical Society},
  issn = {1530-6984},
  doi = {10.1021/acs.nanolett.6b01882},
  urldate = {2025-08-10}
}

@article{Perdew/Burke/Ernzerhof:1996,
  title = {Generalized Gradient Approximation Made Simple},
  author = {Perdew, John P. and Burke, Kieron and Ernzerhof, Matthias},
  year = {1996},
  month = oct,
  journal = {Phys. Rev. Lett.},
  volume = {77},
  number = {18},
  pages = {3865--3868},
  publisher = {American Physical Society},
  doi = {10.1103/PhysRevLett.77.3865},
  urldate = {2022-02-28}
}

@article{Perdew/Yue:1986,
  title = {Accurate and Simple Density Functional for the Electronic Exchange Energy: {{Generalized}} Gradient Approximation},
  author = {Perdew, John P. and Yue, Wang},
  year = {1986},
  month = jun,
  journal = {Phys. Rev. B},
  volume = {33},
  number = {12},
  pages = {8800--8802},
  publisher = {American Physical Society},
  doi = {10.1103/PhysRevB.33.8800},
  urldate = {2025-08-10}
}

@article{Pizzi_et_al:2020,
  title = {Wannier90 as a Community Code: New Features and Applications},
  author = {Pizzi, Giovanni and Vitale, Valerio and Arita, Ryotaro and Bl{\"u}gel, Stefan and Freimuth, Frank and G{\'e}ranton, Guillaume and Gibertini, Marco and Gresch, Dominik and Johnson, Charles and Koretsune, Takashi and {Iba{\~n}ez-Azpiroz}, Julen and Lee, Hyungjun and Lihm, Jae-Mo and Marchand, Daniel and Marrazzo, Antimo and Mokrousov, Yuriy and Mustafa, Jamal I. and Nohara, Yoshiro and Nomura, Yusuke and Paulatto, Lorenzo and Ponc{\'e}, Samuel and Ponweiser, Thomas and Qiao, Junfeng and Th{\"o}le, Florian and Tsirkin, Stepan S. and Wierzbowska, Ma{\l}gorzata and Marzari, Nicola and Vanderbilt, David and Souza, Ivo and Mostofi, Arash A. and Yates, Jonathan R.},
  year = {2020},
  month = jan,
  journal = {J. Phys.: Condens. Matter},
  volume = {32},
  number = {16},
  pages = {165902},
  publisher = {IOP Publishing},
  issn = {0953-8984},
  doi = {10.1088/1361-648X/ab51ff},
  urldate = {2023-04-05}
}

@article{Popescu/Zunger:2012,
  title = {Extracting {{{\emph{E}}}} versus {\emph{k}} Effective Band Structure from Supercell Calculations on Alloys and Impurities},
  author = {Popescu, Voicu and Zunger, Alex},
  year = {2012},
  month = feb,
  journal = {Phys. Rev. B},
  volume = {85},
  number = {8},
  pages = {085201},
  publisher = {American Physical Society},
  doi = {10.1103/PhysRevB.85.085201},
  urldate = {2022-02-24}
}

@article{Pouget_et_al:1974,
  title = {Dimerization of a Linear {{Heisenberg}} Chain in the Insulating Phases of {{V}}{\textsubscript{1-x}}{{Cr}}{\textsubscript{x}}{{O}}{\textsubscript{2}}},
  author = {Pouget, J. P. and Launois, H. and Rice, T. M. and Dernier, P. and Gossard, A. and Villeneuve, G. and Hagenmuller, P.},
  year = {1974},
  month = sep,
  journal = {Phys. Rev. B},
  volume = {10},
  number = {5},
  pages = {1801--1815},
  publisher = {American Physical Society},
  doi = {10.1103/PhysRevB.10.1801},
  urldate = {2022-02-24}
}

@article{Rice/Launois/Pouget:1994,
  title = {Comment on "{{VO}}{\textsubscript{2}}: {{Peierls}} or {{Mott-Hubbard}}? {{A}} View from Band Theory"},
  author = {Rice, T. M. and Launois, H. and Pouget, J. P.},
  year = {1994},
  month = nov,
  journal = {Phys. Rev. Lett.},
  volume = {73},
  number = {22},
  pages = {3042--3042},
  publisher = {American Physical Society},
  doi = {10.1103/PhysRevLett.73.3042},
  urldate = {2023-07-17}
}

@article{Schrecongost_et_al:2019,
  title = {On-{{Demand Nanoscale Manipulations}} of {{Correlated Oxide Phases}}},
  author = {Schrecongost, Dustin and Aziziha, Mina and Zhang, Hai-Tian and Tung, I-Cheng and Tessmer, Joseph and Dai, Weitao and Wang, Qiang and {Engel-Herbert}, Roman and Wen, Haidan and Picard, Yoosuf N. and Cen, Cheng},
  year = {2019},
  journal = {Adv. Funct. Mater.},
  volume = {29},
  number = {49},
  pages = {1905585},
  issn = {1616-3028},
  doi = {10.1002/adfm.201905585},
  urldate = {2025-08-10},
  copyright = {{\copyright} 2019 WILEY-VCH Verlag GmbH \& Co. KGaA, Weinheim}
}

@article{Schrecongost_et_al:2020,
  title = {Rewritable {{Nanoplasmonics}} through {{Room-Temperature Phase Manipulations}} of {{Vanadium Dioxide}}},
  author = {Schrecongost, Dustin and Xiang, Yinxiao and Chen, Jun and Ying, Cuifeng and Zhang, Hai-Tian and Yang, Ming and Gajurel, Prakash and Dai, Weitao and {Engel-Herbert}, Roman and Cen, Cheng},
  year = {2020},
  month = oct,
  journal = {Nano Lett.},
  volume = {20},
  number = {10},
  pages = {7760--7766},
  publisher = {American Chemical Society},
  issn = {1530-6984},
  doi = {10.1021/acs.nanolett.0c03349},
  urldate = {2025-08-10}
}

@article{Schrecongost_et_al:2022,
  title = {Oxygen Vacancy Dynamics in Monoclinic Metallic {{VO}}{\textsubscript{2}} Domain Structures},
  author = {Schrecongost, Dustin and Zhang, Hai-Tian and {Engel-Herbert}, Roman and Cen, Cheng},
  year = {2022},
  month = feb,
  journal = {Appl. Phys. Lett.},
  volume = {120},
  number = {8},
  pages = {081602},
  issn = {0003-6951},
  doi = {10.1063/5.0083771},
  urldate = {2025-08-10}
}

@article{Sim_et_al:2024,
  title = {Crystallographic {{Pathways}} to {{Tailoring Metal-Insulator Transition}} through {{Oxygen Transport}} in {{VO}}{\textsubscript{2}}},
  author = {Sim, Hyeji and Doh, Kyung-Yeon and Park, Yunkyu and Song, Kyung and Kim, Gi-Yeop and Son, Junwoo and Lee, Donghwa and Choi, Si-Young},
  year = {2024},
  journal = {Small},
  volume = {20},
  number = {43},
  pages = {2402260},
  issn = {1613-6829},
  doi = {10.1002/smll.202402260},
  urldate = {2025-08-10},
  copyright = {{\copyright} 2024 The Author(s). Small published by Wiley-VCH GmbH}
}

@article{Souto-Casares/Spaldin/Ederer:2019,
  title = {{{DFT}}+{{DMFT}} Study of Oxygen Vacancies in a {{Mott}} Insulator},
  author = {{Souto-Casares}, Jaime and Spaldin, Nicola A. and Ederer, Claude},
  year = {2019},
  month = aug,
  journal = {Phys. Rev. B},
  volume = {100},
  number = {8},
  pages = {085146},
  issn = {2469-9950, 2469-9969},
  doi = {10.1103/PhysRevB.100.085146},
  urldate = {2023-02-17}
}

@article{Souto-Casares/Spaldin/Ederer:2021,
  title = {Oxygen Vacancies in Strontium Titanate: {{A DFT}}+{{DMFT}} Study},
  author = {{Souto-Casares}, Jaime and Spaldin, Nicola A. and Ederer, Claude},
  year = {2021},
  month = apr,
  journal = {Phys. Rev. Res.},
  volume = {3},
  number = {2},
  pages = {023027},
  issn = {2643-1564},
  doi = {10.1103/PhysRevResearch.3.023027},
  urldate = {2023-02-17}
}

@article{Strelcov_et_al:2012,
  title = {Doping-{{Based Stabilization}} of the {{M2 Phase}} in {{Free-Standing VO}}{\textsubscript{2}} {{Nanostructures}} at {{Room Temperature}}},
  author = {Strelcov, Evgheni and Tselev, Alexander and Ivanov, Ilia and Budai, John D. and Zhang, Jie and Tischler, Jonathan Z. and Kravchenko, Ivan and Kalinin, Sergei V. and Kolmakov, Andrei},
  year = {2012},
  month = dec,
  journal = {Nano Lett.},
  volume = {12},
  number = {12},
  pages = {6198--6205},
  publisher = {American Chemical Society},
  issn = {1530-6984},
  doi = {10.1021/nl303065h},
  urldate = {2023-10-25}
}

@article{Timrov/Marzari/Cococcioni:2018,
  title = {Hubbard Parameters from Density-Functional Perturbation Theory},
  author = {Timrov, Iurii and Marzari, Nicola and Cococcioni, Matteo},
  year = {2018},
  month = aug,
  journal = {Phys. Rev. B},
  volume = {98},
  number = {8},
  pages = {085127},
  publisher = {American Physical Society},
  doi = {10.1103/PhysRevB.98.085127},
  urldate = {2023-07-20}
}

@article{Tomczak/Biermann:2007,
  title = {Effective Band Structure of Correlated Materials: The Case of {{VO}}{\textsubscript{2}}},
  author = {Tomczak, Jan M and Biermann, Silke},
  year = {2007},
  month = sep,
  journal = {J. Phys.: Condens. Matter},
  volume = {19},
  number = {36},
  pages = {365206},
  issn = {0953-8984, 1361-648X},
  doi = {10.1088/0953-8984/19/36/365206},
  urldate = {2022-05-04}
}

@article{Wang_et_al:2016,
  title = {Recent Progress in {{VO}}{\textsubscript{2}} Smart Coatings: {{Strategies}} to Improve the Thermochromic Properties},
  author = {Wang, Shufen and Liu, Minsu and Kong, Lingbing and Long, Yi and Jiang, Xuchuan and Yu, Aibing},
  year = {2016},
  month = aug,
  journal = {Prog. Mater. Sci.},
  volume = {81},
  pages = {1--54},
  publisher = {Elsevier},
  issn = {0079-6425},
  doi = {10.1016/j.pmatsci.2016.03.001},
  urldate = {2022-03-18}
}

@article{Weber_et_al:2012,
  title = {Vanadium Dioxide: {{A Peierls-Mott}} Insulator Stable against Disorder},
  author = {Weber, C{\'e}dric and O'Regan, David D. and Hine, Nicholas D. M. and Payne, Mike C. and Kotliar, Gabriel and Littlewood, Peter B.},
  year = {2012},
  month = jun,
  journal = {Phys. Rev. Lett.},
  volume = {108},
  number = {25},
  pages = {256402},
  issn = {0031-9007, 1079-7114},
  doi = {10.1103/PhysRevLett.108.256402},
  urldate = {2022-04-10}
}

@article{Wegkamp_et_al:2014,
  title = {Instantaneous {{Band Gap Collapse}} in {{Photoexcited Monoclinic VO}}{\textsubscript{2}} Due to {{Photocarrier Doping}}},
  author = {Wegkamp, Daniel and Herzog, Marc and Xian, Lede and Gatti, Matteo and Cudazzo, Pierluigi and McGahan, Christina L. and Marvel, Robert E. and Haglund, Richard F. and Rubio, Angel and Wolf, Martin and St{\"a}hler, Julia},
  year = {2014},
  month = nov,
  journal = {Phys. Rev. Lett.},
  volume = {113},
  number = {21},
  pages = {216401},
  publisher = {American Physical Society},
  doi = {10.1103/PhysRevLett.113.216401},
  urldate = {2025-08-10}
}

@article{Xu_et_al:2016,
  title = {Effects of Annealing Ambient on Oxygen Vacancies and Phase Transition Temperature of {{VO}}{\textsubscript{2}} Thin Films},
  author = {Xu, H. Y. and Huang, Y. H. and Liu, S. and Xu, K. W. and Ma, F. and Chu, Paul K.},
  year = {2016},
  month = aug,
  journal = {RSC Adv.},
  volume = {6},
  number = {83},
  pages = {79383--79388},
  publisher = {The Royal Society of Chemistry},
  issn = {2046-2069},
  doi = {10.1039/C6RA13189A},
  urldate = {2025-08-10}
}

@article{Xue/Yin:2022,
  title = {Element Doping: A Marvelous Strategy for Pioneering the Smart Applications of {{VO}}{\textsubscript{2}}},
  author = {Xue, Yibei and Yin, Shu},
  year = {2022},
  journal = {Nanoscale},
  volume = {14},
  pages = {11054-11097},
  issn = {2040-3364, 2040-3372},
  doi = {10.1039/D2NR01864K},
  urldate = {2022-07-29}
}

@article{Yang/Ko/Ramanathan:2011,
  title = {Oxide Electronics Utilizing Ultrafast Metal-Insulator Transitions},
  author = {Yang, Zheng and Ko, Changhyun and Ramanathan, Shriram},
  year = {2011},
  journal = {Annu. Rev. Mater. Res.},
  volume = {41},
  number = {1},
  pages = {337--367},
  doi = {10.1146/annurev-matsci-062910-100347},
  urldate = {2022-02-28}
}

@article{Yi_et_al:2018,
  title = {Biological Plausibility and Stochasticity in Scalable {{VO}}{\textsubscript{2}} Active Memristor Neurons},
  author = {Yi, Wei and Tsang, Kenneth K. and Lam, Stephen K. and Bai, Xiwei and Crowell, Jack A. and Flores, Elias A.},
  year = {2018},
  month = nov,
  journal = {Nat. Commun.},
  volume = {9},
  pages = {4661},
  issn = {2041-1723},
  doi = {10.1038/s41467-018-07052-w},
  urldate = {2022-06-17},
  pmcid = {PMC6220189},
  pmid = {30405124}
}

@article{Yuan_et_al:2023,
  title = {A Neuromorphic Physiological Signal Processing System Based on {{VO}}{\textsubscript{2}} Memristor for Next-Generation Human-Machine Interface},
  author = {Yuan, Rui and Tiw, Pek Jun and Cai, Lei and Yang, Zhiyu and Liu, Chang and Zhang, Teng and Ge, Chen and Huang, Ru and Yang, Yuchao},
  year = {2023},
  month = jun,
  journal = {Nat. Commun.},
  volume = {14},
  number = {1},
  pages = {3695},
  publisher = {Nature Publishing Group},
  issn = {2041-1723},
  doi = {10.1038/s41467-023-39430-4},
  urldate = {2025-10-10},
  copyright = {2023 The Author(s)}
}

@article{Zhang_et_al:2017,
  title = {Evolution of {{Metallicity}} in {{Vanadium Dioxide}} by {{Creation}} of {{Oxygen Vacancies}}},
  author = {Zhang, Zhen and Zuo, Fan and Wan, Chenghao and Dutta, Aveek and Kim, Jongbum and Rensberg, Jura and Nawrodt, Ronny and Park, Helen Hejin and Larrabee, Thomas J. and Guan, Xiaofei and Zhou, You and Prokes, S. M. and Ronning, Carsten and Shalaev, Vladimir M. and Boltasseva, Alexandra and Kats, Mikhail A. and Ramanathan, Shriram},
  year = {2017},
  month = mar,
  journal = {Phys. Rev. Appl.},
  volume = {7},
  number = {3},
  pages = {034008},
  publisher = {American Physical Society},
  doi = {10.1103/PhysRevApplied.7.034008},
  urldate = {2024-10-03}
}

@article{Zhang_et_al:2017a,
  title = {Evolution of {{Structural}} and {{Electrical Properties}} of {{Oxygen-Deficient VO}}{\textsubscript{2}} under {{Low Temperature Heating Process}}},
  author = {Zhang, Jiasong and Zhao, Zhengjing and Li, Jingbo and Jin, Haibo and Rehman, Fida and Chen, Pengwan and Jiang, Yijie and Chen, Chunxu and Cao, Maosheng and Zhao, Yongjie},
  year = {2017},
  month = aug,
  journal = {ACS Appl. Mater. Interfaces},
  volume = {9},
  number = {32},
  pages = {27135--27141},
  publisher = {American Chemical Society},
  issn = {1944-8244},
  doi = {10.1021/acsami.7b05792},
  urldate = {2025-08-10}
}

\end{document}